\documentclass[onecolumn]{emulateapj}

\shorttitle{Estimating Luminosity Functions}
\shortauthors{Kelly et al.}

\begin{document}
  
  \title{A Flexible Method of Estimating Luminosity Functions}

  \author{Brandon C. Kelly\altaffilmark{1}, Xiaohui Fan,}
  \affil{Steward Observatory, University of Arizona, 933 N Cherry Ave,
  Tucson, AZ 85721}
  \and
  \author{Marianne Vestergaard}
  \affil{Dept. of Physics and Astronomy, Robinson Hall, Tufts University, 
    Medford, MA 02155}
  \altaffiltext{1}{bkelly@as.arizona.edu}

  \begin{abstract}
    We describe a Bayesian approach to estimating luminosity
    functions. We derive the likelihood function and posterior
    probability distribution for the luminosity function, given the
    observed data, and we compare the Bayesian approach with
    maximum-likelihood by simulating sources from a Schechter
    function. For our simulations confidence intervals derived from
    bootstrapping the maximum-likelihood estimate can be too narrow,
    while confidence intervals derived from the Bayesian approach are
    valid. We develop our statistical approach for a flexible model
    where the luminosity function is modeled as a mixture of Gaussian
    functions. Statistical inference is performed using Markov chain
    Monte Carlo (MCMC) methods, and we describe a Metropolis-Hastings
    algorithm to perform the MCMC. The MCMC simulates random draws
    from the probability distribution of the luminosity function
    parameters, given the data, and we use a simulated data set to
    show how these random draws may be used to estimate the
    probability distribution for the luminosity function. In addition,
    we show how the MCMC output may be used to estimate the
    probability distribution of any quantities derived from the
    luminosity function, such as the peak in the space density of
    quasars. The Bayesian method we develop has the advantage that it
    is able to place accurate constraints on the luminosity function
    even beyond the survey detection limits, and that it provides a
    natural way of estimating the probability distribution of any
    quantities derived from the luminosity function, including those
    that rely on information beyond the survey detection limits.
  \end{abstract}
  
  
  \section{INTRODUCTION}

  The luminosity function (LF) has been an important tool for
  understanding the evolution of galaxies and quasars, as it provides
  a census of the galaxy and quasar populations over cosmic
  time. Quasar luminosity functions have been estimated for optical
  surveys \citep[e.g.,][]{fan01,wolf03,croom04,dr3lumfunc,jiang06},
  X-ray surveys \citep[e.g.,][]{steffen03,ueda03,barger05,lafranca05},
  infrared surveys \citep[e.g.,][]{barger05,matute06,babb06}, radio
  surveys \citep[e.g.,][]{wadd01,will01}, and emission lines
  \citep{hao05}. In addition, luminosity functions across different
  bands have been combined to form an estimate of the bolometric
  luminosity function \citep{hopkins07}. Besides providing an
  important constraint on models of quasar evolution and supermassive
  black hole growth \citep[e.g.,][]{wyithe03,hopkins06}, studies of
  the LF have found evidence for `cosmic downsizing', where the space
  density of more luminous quasars peaks at higher redshift. Attempts
  to map the growth of supermassive black holes start from the local
  supermassive black hole distribution, and employ the argument of
  \citet{soltan82}, using the quasar luminosity function as a
  constraint on the black hole mass distribution. These studies have
  found evidence that the highest mass black holes grow first
  \citep[e.g.,][]{yu02,marconi04,merloni04}, suggesting that this
  cosmic downsizing is the result of an anti-hierarchical growth of
  supermassive black holes.

  Similarly, galaxy luminosity functions have been estimated in the
  optical \citep[e.g.,][]{blanton03,dahlen05,brown07,march07a}, X-ray
  \citep[e.g.,][]{kim06,ptak07}, infrared
  \citep[e.g.,][]{ciras07,huynh07}, ultraviolet
  \citep[e.g.,][]{bud05,paltani07}, radio
  \citep[e.g.,][]{lin07,mauch07}, for galaxies in clusters
  \citep[e.g.,][]{popess06,harsono07}, and for galaxies in voids
  \citep{hoyle05}. The galaxy luminosity function probes several
  aspects of the galaxy population; namely (a) the evolution of
  stellar populations and star formation histories
  \citep[e.g.,][]{faber07}, (b) the local supermassive black hole mass
  distribution \citep[e.g,][]{yu02,marconi04} via the Magorrian
  relationship \citep{mag98}, (c) the dependence of galaxy properties
  on environment \citep[e.g.,][]{croton05,lauer07}, and (d) places
  constraints on models of structure formation and galaxy evolution
  \citep[e.g.,][]{bower06,finlator06,march07b}.

  Given the importance of the luminosity function as an observational
  constraint on models of quasar and galaxy evolution, it is essential
  that a statistically accurate approach be employed when estimating
  these quantities. However, the existence of complicated selection
  functions hinders this, and, as a result, a variety of methods have
  been used to accurately account for the selection function when
  estimating the LF. These include various binning methods
  \citep[e.g.,][]{schmidt68,avni80,page00}, maximum-likelihood fitting
  \citep[e.g.,][]{marshall83,fan01}, and a powerful semi-parametric
  approach \citep{schafer07}. In addition, there have been a variety
  of methods proposed for estimating the cumulative distribution
  function of the LF \citep[e.g.,][]{lynden71,efron92,mal99}.

  Each of these statistical methods has advantages and
  disadvantages. Statistical inference based on the binning procedures
  cannot be extended beyond the support of the selection function, and
  the cumulative distribution function methods typically assume that
  luminosity and redshift are statistically independent. Furthermore,
  one is faced with the arbitrary choice of bin size. The
  maximum-likelihood approach typically assumes a restrictive and
  somewhat \emph{ad hoc} parametric form, and has not been used to
  give an estimate of the LF normalization; instead, for example, the
  LF normalization is often chosen to make the expected number of
  sources detected in one's survey equal to the actual number of
  sources detected. In addition, confidence intervals based on the
  errors derived from the various procedures are typically derived by
  assuming that the uncertainties on the LF parameters have a Gaussian
  distribution. While this is valid as the sample size approaches
  infinity, it is not necessarily a good approximation for finite
  sample sizes. This is particularly problematic if one is employing
  the best fit results to extrapolating the luminosity function beyond
  the bounds of the selection function. It is unclear if the
  probability distribution of the uncertainty in the estimated
  luminosity function below the flux limit is even asymptotically
  normal.

  Motivated by these issues, we have developed a Bayesian method for
  estimating the luminosity function. We derive the likelihood
  function of the LF by relating the observed data to the true LF,
  assuming some parametric form, and derive the posterior probability
  distribution of the LF parameters, given the observed data. While
  the likelihood function and posterior are valid for any parametric
  form, we focus on a flexible parametric model where the LF is
  modeled as a weighted sum of Gaussian functions. This is a type of
  `non-parametric' approach, where the basic idea is that the
  individual Gaussian functions do not have any physical meaning, but
  that given enough Gaussian functions one can obtain a suitably
  accurate approximation to the true LF; a similar approach has been
  taken by \citet{blanton03} for estimating galaxy LFs, and by
  \citet{kelly07} within the context of linear regression with
  measurement error. Modeling the LF as a mixture of Gaussian
  functions avoids the problem of choosing a particular parametric
  form, especially in the absence of any guidance from astrophysical
  theory. The mixture of Gaussians model has been studied from a
  Bayesian perspective by numerous authors
  \citep[e.g.,][]{roeder97,jasra05,dell06}. In addition, we describe a
  Markov chain Monte Carlo (MCMC) algorithm for obtaining random draws
  from the posterior distribution. These random draws allow one to
  estimate the posterior distribution for the LF, as well as any
  quantities derived from it. The MCMC method therefore allows a
  straight-forward method of calculating uncertainties on any quantity
  derived from the LF, such as the redshift where the space density of
  quasars or galaxies peaks; this has proven to be a challenge for
  other statistical methods developed for LF estimation. Because the
  Bayesian approach is valid for any sample size, one is therefore
  able to place reliable constraints on the LF and related quantities
  even below the survey flux limits.

  Because of the diversity and mathematical complexity of some parts
  of this paper, we summarize the main results here. We do this so
  that the reader who is only interested in specific aspects of this
  paper can conveniently consult the sections of interest.
  \begin{itemize}
  \item
    In \S~\ref{s-lik} we derive the general form of the likelihood
    function for luminosity function estimation. We show that the
    commonly used likelihood function based on the Poisson
    distribution is incorrect, and that the correct form of the
    likelihood function is derived from the binomial
    distribution. However, because the Poisson distribution is the
    limit of the binomial distribution as the probability of including
    a source in a survey approaches zero, the maximum-likelihood
    estimates derived from the two distribution give nearly identical
    results so long as a survey's detection probability is small. The
    reader who is interested in using the correct form of the
    likelihood function of the LF should consult this section.
  \item
    In \S~\ref{s-posterior} we describe a Bayesian approach to
    luminosity function estimation. We build on the likelihood
    function derived in \S~\ref{s-lik} to derive the probability
    distribution of the luminosity function, given the observed data
    (i.e., the posterior distribution). We use a simple example based
    on a Schechter function to illustrate the Bayesian approach, and
    compare it with the maximum-likelihood approach. For this example,
    we find that confidence intervals derived from the posterior
    distribution are valid, while confidence intervals derived from
    bootstrapping the maximum-likelihood estimate can be too
    small. The reader who is interested in a Bayesian approach to
    luminosity function estimation, and how it compares with
    maximum-likelihood, should consult this section.
  \item
    In \S~\ref{s-smodel} we develop a mixture of Gaussian functions
    model for the luminosity function, deriving the likelihood
    function and posterior distribution for the model. Under this
    model, the LF is modeled as a weighted sum of Gaussian
    functions. This model has the advantage that given a suitably
    large enough number of Gaussian functions, it is flexible enough
    to give an accurate estimate of any smooth and continuous LF. This
    allows the model to adapt to the true LF, thus minimizing the bias
    that can result when assuming a parametric form of the LF. This
    is particularly useful when extrapolating beyond the flux limits
    of a survey, where bias caused by parametric misspecification can
    be a significant concern. The reader who are interested in
    employing the mixture of Gaussian functions model should consult
    this section.
  \item
    Because of the large number of parameters often associated with
    luminosity function estimation, Bayesian inference is most easily
    performed by obtaining random draws of the LF from the posterior
    distribution.  In \S~\ref{s-mha} we describe the
    Metropolis-Hastings algorithm (MHA) for obtaining random draws of
    the LF from the posterior distribution. As an example, we describe
    a MHA for obtaining random draws of the parameters for a Schechter
    function from the posterior distribution. Then, we describe a more
    complex MHA for obtaining random draws of the parameters for the
    mixture of Gaussian functions model. The reader who is interested
    in the computational aspects of `fitting' the mixture of Gaussian
    functions model, or who is interested in the computational aspects
    of Bayesian inference for the LF, should consult this section. A
    computer routine for performing the Metropolis-Hastings algorithm
    for the mixture of Gaussian functions model is available on
    request from B. Kelly.
  \item
    In \S~\ref{s-sim} we use simulation to illustrate the
    effectiveness of our Bayesian Gaussian mixture model for
    luminosity function estimation. We construct a simulated data set
    similar to the Sloan Digital Sky Survey DR3 Quasar Catalog
    \citep{dr3qsos}. We then use our mixture of Gaussian functions
    model to recover the true LF and show that our mixture model is
    able to place reliable constraints on the LF. We also illustrate
    how to use the MHA output to constrain any quantity derived from
    the LF, and how to use the MHA output to assess the quality of the
    fit. The reader who is interested in assessing the effectiveness
    of our statistical approach, or who is interested in using the MHA
    output for statistical inference on the LF, should consult this
    section.
  \end{itemize}
  We adopt a cosmology based on the the WMAP best-fit parameters
  \citep[$h=0.71, \Omega_m=0.27, \Omega_{\Lambda}=0.73$,][]{wmap}

  \section{THE LIKELIHOOD FUNCTION}

  \label{s-lik}

  \subsection{Notation}

  \label{s-notation}

  We use the common statistical notation that an estimate of a
  quantity is denoted by placing a `hat' above it; e.g.,
  $\hat{\theta}$ is an estimate of the true value of the parameter
  $\theta$. The parameter $\theta$ may be scalar or multivalued. We
  denote a normal density\footnote[1]{We use the terms probability
  density and probability distribution interchangeably.}  (i.e., a
  Gaussian distribution) with mean $\mu$ and variance $\sigma^2$ as
  $N(\mu, \sigma^2)$, and we denote as $N_p (\mu, \Sigma)$ a
  multivariate normal density with $p$-element mean vector $\mu$ and
  $p \times p$ covariance matrix $\Sigma$. If we want to explicitly
  identify the argument of the Gaussian function, we use the notation
  $N(x|\mu,\sigma^2)$, which should be understood to be a Gaussian
  with mean $\mu$ and variance $\sigma^2$ as a function of $x$. We
  will often use the common statistical notation where ``$\sim$''
  means ``is drawn from'' or ``is distributed as''. This should not be
  confused with the common usage of implying ``similar to''. For
  example, $x \sim N(\mu, \sigma^2)$ states that $x$ is drawn from a
  normal density with mean $\mu$ and variance $\sigma^2$, whereas $x
  \sim 1$ states that the value of $x$ is similar to one.

  In this work, the maximum-likelihood estimate of the luminosity
  function refers to an estimate of the LF obtained by maximizing the
  likelihood function of the unbinned data. Therefore, the
  maximum-likelihood estimate does not refer to an estimate obtained
  by maximizing the likelihood function of binned data, such as
  fitting the results obtained from the $1 / V_a$ technique.

  \subsection{Derivation of the Luminosity Function Likelihood}

  \label{s-lflik}

  The luminosity function, denoted as $\phi(L, z) dL$, is the number
  of sources per comoving volume $V(z)$ with luminosities in the range
  $L, L + dL$. The luminosity function is related to the probability
  density of $(L,z)$ by
  \begin{equation}
    p(L,z) = \frac{1}{N} \phi(L,z) \frac{dV}{dz},
    \label{eq-phiconvert}
  \end{equation}
  where $N$ is the total number of sources in the observable universe,
  and is given by the integral of $\phi$ over $L$ and $V(z)$. Note
  that $p(L,z) dL dz$ is the probability of finding a source in the
  range $L, L + dL$ and $z, z + dz$. Equation (\ref{eq-phiconvert})
  separates the LF into its shape, given by $p(L,z)$, and its
  normalization, given by $N$. Once we have an estimate of $p(L,z)$,
  we can easily convert this to an estimate of $\phi(L,z)$ using
  Equation (\ref{eq-phiconvert}). In general, it is easier to work
  with the probability distribution of $L$ and $z$, instead of
  directly with the LF, because $p(L,z)$ is more directly related to
  the likelihood function.

  If we assume a parametric form for $\phi(L,z)$, with parameters
  $\theta$, we can derive the likelihood function for the observed
  data. The likelihood function is the probability of observing one's
  data, given the assumed model. The presence of flux limits and
  various other selection effects can make this difficult, as the
  observed data likelihood function is not simply given by Equation
  (\ref{eq-phiconvert}). In this case, the set of luminosities and
  redshifts observed by a survey gives a biased estimate of the true
  underlying distribution, since only those sources with $L$ above the
  flux limit at a given $z$ are detected. In order to derive the
  observed data likelihood function, it is necessary to take the
  survey's selection method into account. This is done by first
  deriving the joint likelihood function of both the observed and
  unobserved data, and then integrating out the unobserved data.

  Because the data points are independent, the likelihood function for
  all $N$ sources in the universe is
  \begin{equation}
    p(L,z|\theta) = \prod_{i=1}^{N} p(L_i,z_i|\theta) \label{eq-fulllik}.
  \end{equation}
  In reality, we do not know the luminosities and redshifts for all
  $N$ sources, nor do we know the value of $N$, as our survey only
  covers a fraction of the sky and is subject to a selection
  function. As a result, our survey only contains $n$ sources. Because
  of this, the selection process must also be included in the
  probability model, and the total number of sources, $N$, is an
  additional parameter that needs to be estimated.

  We can incorporate the sample selection into the likelihood function
  by including the random detection of sources. We introduce an
  $N$-element indicator vector ${\bf I}$ that takes on the values $I_i
  = 1$ if the $i^{\rm th}$ source is included in our survey and $I_i =
  0$ otherwise. Note that ${\bf I}$ is a vector of size $N$ containing
  only ones and zeros. In this case, the selection function is the
  probability of including a source given $L$ and $z$, $p(I_i =
  1|L_i,z_i)$. The complete data likelihood is then the probability
  that all objects of interest in the universe (e.g., all quasars)
  have luminosities $L_1,\ldots,L_N$ and redshifts $z_1,\ldots,z_N$,
  and that the selection vector $I$ has the values $I_1,\ldots,I_N$,
  given our assumed luminosity function:
  \begin{equation} 
    p(L,z,{\bf I}|\theta,N) = C^N_n \prod_{i \in {\cal A}_{obs}} 
    p(I_i = 1|L_i, z_i) p(L_i, z_i | \theta) \prod_{j \in {\cal A}_{mis}} 
    p(I_j = 0|L_j, z_j) p(L_j, z_j|\theta).
    \label{eq-complik1}
  \end{equation}
  Here, $C^N_n = N! / n! (N - n)!$ is the binomial coefficient, ${\cal
  A}_{obs}$ denotes the set of $n$ included sources, and ${\cal
  A}_{mis}$ denotes the set of $N - n$ missing sources. The number of
  sources detected in a survey is random, and therefore the binomial
  coefficient is necessary in normalizing the likelihood function, as
  it gives the number of possible ways to select a subset of $n$
  sources from a set of $N$ total sources.

  Because we are interested in the probability of the observed data,
  given our assumed model, the complete data likelihood function is of
  little use by itself. However, we can integrate Equation
  (\ref{eq-complik1}) over the missing data to obtain the observed
  data likelihood function. This is because the marginal probability
  distribution of the observed data is obtained by integrating the
  joint probability distribution of the observed and the missing data
  over the missing data:
  \begin{eqnarray}
    p(L_{obs},z_{obs},{\bf I}|\theta,N) & = & C^N_n \prod_{i \in {\cal A}_{obs}} 
    p(I_i = 1|L_i, z_i) p(L_i, z_i | \theta)  \\
    & \times & \prod_{j \in {\cal A}_{mis}} \int_{0}^{\infty} \int_{0}^{\infty} 
    p(I_j = 0|L_j, z_j) p(L_j, z_j|\theta)\ dL_j\ dz_j \label{eq-obslik0} \\
    & \propto & C^N_n \left[p(I=0|\theta) \right]^{N-n} 
    \prod_{i \in {\cal A}_{obs}} p(L_i,z_i | \theta),
    \label{eq-obslik1}
  \end{eqnarray}
  where the probability that the survey misses a source, given the
  parameters $\theta$, is
  \begin{equation}
    p(I=0|\theta) = \int \int p(I = 0|L,z) p(L,z|\theta)\ dL\ dz.
    \label{eq-selprob}
  \end{equation}
  Here, we have introduced the notation that $L_{obs}$ and $z_{obs}$
  denote the set of values of $L$ and $z$ for those sources included
  in one's survey, and we have omitted terms that do not depend on
  $\theta$ or $N$ from Equation (\ref{eq-obslik1}). Equation
  (\ref{eq-obslik1}) is the observed data likelihood function, given
  an assumed luminosity function
  (Eq.[\ref{eq-phiconvert}]). Qualitatively, the observed data
  likelihood function is the probability of observing the set of $n$
  luminosities $L_1, \ldots, L_n$ and redshifts $z_1, \ldots, z_n$
  given the assumed luminosity function parameterized by $\theta$,
  multiplied by the probability of not detecting $N - n$ sources given
  $\theta$, multiplied by the number of ways of selecting a subset of
  $n$ sources from a set of $N$ total sources. The observed data
  likelihood function can be used to calculate a maximum likelihood
  estimate of the luminosity function, or combined with a prior
  distribution to perform Bayesian inference.

  \subsection{Comparison with the Poisson Likelihood}

  \label{s-lumfunc}

  The observed data likelihood given by Equation (\ref{eq-obslik1})
  differs from that commonly used in the luminosity function
  literature. Instead, a likelihood based on the Poisson distribution
  is often used. \citet{marshall83} give the following equation for
  the log-likelihood function based on the Poisson distribution:
  \begin{equation}
    \log p(L_{obs}, z_{obs}|\theta,N) = \sum_{i \in {\cal A}_{obs}} \log
    \phi(L_i, z_i |N, \theta) + \int \int p(I=1|L,z) \phi(L,z|\theta,N) \frac{dV}{dz}\ dL\ dz.
    \label{eq-poislik1}
  \end{equation}
  Inserting Equation (\ref{eq-phiconvert}) for $\phi(L,z|\theta)$, the
  log-likelihood based on the Poisson likelihood becomes
  \begin{equation}
    \log p(L_{obs}, z_{obs}|\theta,N) = n \log N + \sum_{i \in {\cal A}_{obs}} \log
    p(L_i, z_i |\theta) - N p(I=1|\theta),
    \label{eq-poislik2}
  \end{equation}
  where, $p(I=1|\theta) = 1 - p(I=0|\theta)$, and $p(I=0|\theta)$ is
  given by Equation (\ref{eq-selprob}). In contrast, the
  log-likelihood we have derived based on the binomial distribution is
  the logarithm of Equation (\ref{eq-obslik1}):
  \begin{equation}
    \log p(L_{obs}, z_{obs}|\theta,N) = \log N! - \log n! - \log (N-n)! + 
    \sum_{i \in {\cal A}_{obs}} \log p(L_i, z_i |\theta) + (N-n) \log p(I=0|\theta).
    \label{eq-loglik}
  \end{equation}
  The likelihood functions implied by Equations (\ref{eq-poislik2})
  and (\ref{eq-loglik}) are functions of $N$, and thus the likelihoods
  may also be maximized with respect to the LF normalization. This is
  contrary to what is often claimed in the literature, where the LF
  normalization is typically chosen to make the expected number of
  sources observed in one's survey equal to the actual number
  observed.

  The binomial likelihood, given by Equation (\ref{eq-obslik1}),
  contains the term $C^N_n$, resulting from the fact that the total
  number of sources included in a survey, $n$, follows a binomial
  distribution. For example, suppose one performed a survey over one
  quarter of the sky with no flux limit. Assuming that sources are
  uniformly distributed on the sky, the probability of including a
  source for this survey is simply $1/4$. If there are $N$ total
  sources in the universe, the total number of sources that one would
  find within the survey area follows a binomial distribution with $N$
  `trials' and probability of `success' $p = 1/4$. However, the
  Poisson likelihood is derived by noting that the number of sources
  detected in some small bin in $(L,z)$ follows a Poisson
  distribution. Since the sum of a set of Poisson distributed random
  variables also follows a Poisson distribution, this implies that the
  total number of sources detected in one's survey, $n$, follows a
  Poisson distribution. However, $n$ actually follows a binomial
  distribution, and thus the observed data likelihood function is not
  given by the Poisson distribution. The source of this error is
  largely the result of approximating the number of sources in a bin
  as following a Poisson distribution, when in reality it follows a
  binomial distribution.

  Although the Poisson likelihood function for the LF is incorrect,
  the previous discussion should not be taken as a claim that previous
  work based on the Poisson likelihood function is incorrect. When the
  number of sources included in one's sample is much smaller than the
  total number of sources in the universe, the binomial distribution
  is well approximated by the Poisson distribution. Therefore, if the
  survey only covers a small fraction of the sky, or if the flux limit
  is shallow enough such that $n \ll N$, then the Poisson likelihood
  function should provide an accurate approximation to the true
  binomial likelihood function. When this is true, statistical
  inference based on the Poisson likelihood should only exhibit
  negligible error, so long as there are enough sources in one's
  survey to obtain an accurate estimate of the LF normalization. In
  \S~\ref{s-schechter2} we use simulate to compare results obtained
  from the two likelihood functions, and to compare the
  maximum-likelihood approach to the Bayesian approach.

  \section{POSTERIOR DISTRIBUTION FOR THE LF PARAMETERS}

  \label{s-posterior}

  We can combine the likelihood function for the LF with a prior
  probability distribution on the LF parameters to perform Bayesian
  inference on the LF. The result is the posterior probability
  distribution of the LF parameters, i.e., the probability
  distribution of the LF parameters given our observed data. This is
  in contrast to the maximum likelihood approach, where the maximum
  likelihood approach seeks to relate the observed value of the MLE to
  the true parameter value through an estimate of the sampling
  distribution of the MLE. In Appendix \S~\ref{a-mle_vs_bayes} we give
  a more thorough introduction to the difference between the maximum
  likelihood and Bayesian approaches.

  \subsection{Derivation of the Posterior Probability Distribution}

  \label{s-postderiv}

  The posterior probability distribution of the model parameters is
  related to the likelihood function and the prior probability
  distribution as
  \begin{equation}
    p(\theta, N|L_{obs},z_{obs},{\bf I}) \propto p(\theta, N)
    p(L_{obs},z_{obs},{\bf I} | \theta, N),
    \label{eq-post}
  \end{equation}
  where $p(\theta, N)$ is the prior on $(\theta, N)$, and $p(L_{obs},
  z_{obs}, {\bf I}| \theta, N)$ is is the observed data likelihood
  function, given by Equation (\ref{eq-obslik1}). The posterior
  distribution is the probability distribution of $\theta$ and $N$,
  given the observed data, $L_{obs}$ and $z_{obs}$. Because the
  luminosity function depends on the parameters $\theta$ and $N$, the
  posterior distribution of $\theta$ and $N$ can be used to obtain the
  probability distribution of $\phi(L,z)$, given our observed set of
  luminosities and redshifts.

  It is of use to decompose the posterior as
  $p(\theta,N|L_{obs},z_{obs}) \propto p(N|\theta,L_{obs},z_{obs})
  p(\theta|L_{obs},z_{obs})$; here we have dropped the explicit
  conditioning on ${\bf I}$. This decomposition separates the
  posterior into the conditional posterior of the LF normalization at
  a given $\theta$, $p(N|L_{obs},z_{obs},\theta)$, from the marginal
  posterior of the LF shape, $p(\theta|L_{obs},z_{obs})$. In this work
  we assume that $N$ and $\theta$ are independent in their prior
  distribution, $p(\theta,N) = p(N) p(\theta)$, and that the prior on
  $N$ is uniform over $\log N$. A uniform prior on $\log N$
  corresponds to a prior distribution on $N$ of $p(N) \propto 1/N$, as
  $p(\log N) d \log N = p(N) dN$. Under this prior, one can show that
  the marginal posterior probability distribution of $\theta$ is
  \begin{equation}
    p(\theta|L_{obs},z_{obs}) \propto p(\theta) \left[p(I=1|\theta) \right]^{-n} 
    \prod_{i \in {\cal A}_{obs}} p(L_i,z_i|\theta),
    \label{eq-thetapost}
  \end{equation}
  where $p(I=1|\theta) = 1 - p(I=0|\theta)$. We derive Equation
  (\ref{eq-thetapost}) in Appendix \S~\ref{a-margpost_deriv}
  \citep[see also][]{gelman04}. Under the assumption of a uniform
  prior on $\theta$, Equation (\ref{eq-thetapost}) is equivalent to
  Equation (22) in \citet{fan01}, who use a different derivation to
  arrive at a similar result.

  Under the prior $p(\log N) \propto 1$, the conditional posterior
  distribution of $N$ at a given $\theta$ is a negative binomial
  distribution with parameters $n$ and $p(I=1|\theta)$. The negative
  binomial distribution gives the probability that the total number of
  sources in the universe is equal to $N$, given that we have observed
  $n$ sources in our sample with probability of inclusion
  $p(I=1|\theta)$:
  \begin{equation}
    p(N|n,\theta) = C^{N-1}_{n-1} \left[p(I=1|\theta)\right]^n 
    \left[p(I=0|\theta)\right]^{N-n}.
    \label{eq-npost}
  \end{equation}
  Here, $p(I=0|\theta)$ is given by Equation (\ref{eq-selprob}) and
  $p(I=1|\theta) = 1 - p(I=0|\theta)$. Further description of the
  negative binomial distribution is given in \S~\ref{a-densities}. The
  complete joint posterior distribution of $\theta$ and $N$ is then
  the product of Equations (\ref{eq-thetapost}) and (\ref{eq-npost}),
  $p(\theta,N|L_{obs},z_{obs}) \propto p(N|\theta,n)
  p(\theta|L_{obs},z_{obs})$.

  Because it is common to fit a luminosity function with a large
  number of parameters, it is computationally intractable to directly
  calculate the posterior distribution from Equations
  (\ref{eq-thetapost}) and (\ref{eq-npost}). In particular, the number
  of grid points needed to calculate the posterior will scale
  exponentially with the number of parameters. Similarly, the number
  of integrals needed to calculate the marginal posterior probability
  distribution of a single parameters will also increase exponentially
  with the number of parameters. Instead, Bayesian inference is most
  easily performed by simulating random draws of $N$ and $\theta$ from
  their posterior probability distribution. Based on the decomposition
  $p(\theta,N|L_{obs},z_{obs}) \propto p(N|n,\theta)
  p(\theta|L_{obs},z_{obs})$, we can obtain random draws of
  $(\theta,N)$ from the posterior by first drawing values of $\theta$
  from Equation (\ref{eq-thetapost}). Then, for each draw of $\theta$,
  we draw a value of $N$ from the negative binomial distribution. The
  values of $N$ and $\theta$ can then be used to compute the values of
  luminosity function via Equation (\ref{eq-phiconvert}). The values
  of the LF computed from the random draws of $N$ and $\theta$ are
  then treated as a random draw from the probability distribution of
  the LF, given the observed data. These random draws can be used to
  estimate posterior means and variances, confidence intervals, and
  histogram estimates of the marginal distributions. Random draws for
  $\theta$ may be obtained via Markov chain Monte Carlo (MCMC)
  methods, described in \S~\ref{s-mha}, and we describe in
  \S~\ref{a-densities} how to obtain random draws from the negative
  binomial distribution. In \S~\ref{s-simmcmc} we give more details on
  using random draws from the posterior to perform statistical
  inference on the LF.

  \subsection{Illustration of the Bayesian Approach: Schechter Function}

  \label{s-schechter}

  Before moving to more advanced models, we illustrate the Bayesian
  approach by applying it to a simulated set of luminosities drawn
  from a Schechter function. We do this to give an example of how to
  calculate the posterior distribution, how to obtain random draws
  from the posterior and use these random draws to draw scientific
  conclusions based on the data, and to compare the Bayesian approach
  with the maximum-likelihood approach (see
  \S~\ref{s-schechter2}). The Schechter luminosity function is:
  \begin{equation}
    \phi(L) = \frac{N}{L^{*} \Gamma(\alpha + 1)} \left( \frac{L}{L^{*}} \right)^{\alpha}
    e^{-L / L^{*}}, \ \ \theta = (\alpha, L^{*}).
    \label{eq-schechter}
  \end{equation}
  For simplicity, we ignore a $z$ dependence. The Schechter function
  is equivalent to a Gamma distribution with shape parameter $k =
  \alpha + 1$, and scale parameter $L^*$. Note that $k > 0$ and
  $\alpha > -1$; otherwise the integral of Equation
  (\ref{eq-schechter}) may be negative or become infinite. For our
  simulation, we randomly draw $N = 1000$ galaxy luminosities from
  Equation (\ref{eq-schechter}) using a value of $\alpha = 0$ and $L^*
  = 10^{44}\ {\rm erg\ s^{-1}}$.

  To illustrate how the results depend on the detection limit, we
  placed two different detection limits on our simulated survey. The
  first limit was at $L_{min} = 2 \times 10^{43}\ {\rm ergs\ s^{-1}},$
  and the second was at $L_{min} = 2 \times 10^{44}\ {\rm ergs\
  s^{-1}}$. We used a hard detection limit, where all sources above
  $L_{min}$ were detected and all sources below $L_{min}$ were not:
  $p(I=1|L > L_{min}) = 1$ and $p(I=1|L < L_{min}) = 0$. Note that the
  first detection limit lies below $L^*$, while the second detection
  limit lies above $L^*$. We were able to detect $n \sim 818$ sources
  for $L_{min} = 2 \times 10^{43}\ {\rm ergs\ s^{-1}}$ and $n \sim
  135$ sources for $L_{min} = 2 \times 10^{44}\ {\rm ergs\ s^{-1}}$

  The marginal posterior distribution of $\alpha$ and $L^*$ can be
  calculated by inserting into Equation (\ref{eq-thetapost}) an
  assumed prior probability distribution, $p(\alpha, L^*)$, and the
  likelihood function, $p(L_i|\alpha, L^*)$. Because we are ignoring
  redshift in our example, the likelihood function is simply
  $p(L_i|\alpha,L^*) = \phi(L) / N$. In this example, we assume a
  uniform prior on $\log L^*$ and $\alpha$, and therefore $p(L^*,
  \alpha) \propto 1 / L^*$. From Equations (\ref{eq-thetapost}) and
  (\ref{eq-schechter}), the marginal posterior distribution of the
  parameters is
  \begin{equation}
    p(\alpha, L^*|L_{obs}) \propto \frac{1}{L^*} \left[ p(I=1|\alpha, L^*) \right]^{-n} 
    \prod_{i=1}^n \frac{1}{L^* \Gamma(\alpha + 1)} \left( \frac{L_i}{L^*} \right)^{\alpha}
    e^{-L_i / L^*},
    \label{eq-schechpost}
  \end{equation}
  where the survey detection probability is
  \begin{equation}
    p(I=1|\alpha, L^*) = \int_{L_{min}}^{\infty} \frac{1}{L^* \Gamma(\alpha + 1)} 
    \left( \frac{L_i}{L^*} \right)^{\alpha} e^{-L_i / L^*}\ dL.
    \label{eq-schechdet}
  \end{equation}
  The conditional posterior distribution of $N$ at a given $\theta$ is
  given by inserting in Equation (\ref{eq-schechdet}) into Equation
  (\ref{eq-npost}), and the joint posterior of $\alpha, L^*,$ and $N$
  is obtained by multiplying Equation (\ref{eq-schechpost}) by
  Equation (\ref{eq-npost}).

  We perform statistical inference on the LF by obtaining random draws
  from the posterior distribution. In order to calculate the marginal
  posterior distributions, $p(\alpha|L_{obs}), p(L^*|L_{obs}),$ and
  $p(N|L_{obs})$, we would need to numerically integrate the posterior
  distribution over the other two parameters. For example, in order to
  calculate the marginal posterior of $\alpha$,
  $p(\alpha|L_{obs},z_{obs})$, we would need to integrate
  $p(\alpha,L^*,N|L_{obs})$ over $L^*$ and $N$ on a grid of values for
  $\alpha$. While feasible for the simple 3-dimensional problem
  illustrated here, it is faster to simply obtain a random draw of
  $\alpha,L^*,$ and $N$ from the posterior, and then use a histogram
  to estimate $p(\alpha|L_{obs})$. Further details are given in
  \S~\ref{s-simmcmc} on performing Bayesian inference using random
  draws from the posterior.

\begin{figure}
    \begin{center}
      \includegraphics[scale=0.5, angle=90]{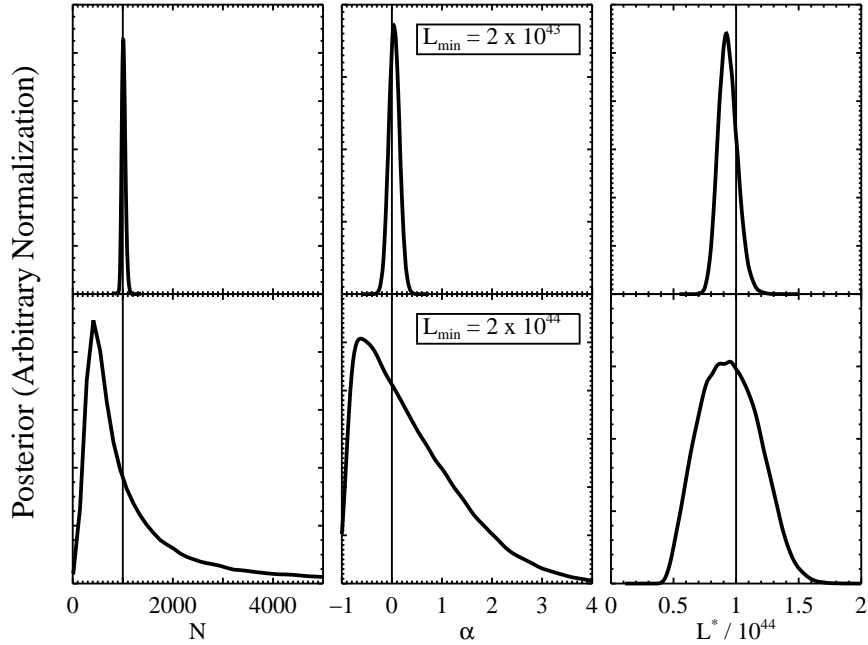}
      \caption{Posterior probability distribution of the Schechter
	luminosity function parameters, $N, \alpha,$ and $L^*$, for the
	simulated sample described in \S~\ref{s-schechter}. The top three
	panels show the posterior when the luminosity limit of the survey
	is $L > 2 \times 10^{43}\ [{\rm erg\ s^{-1}}]$, and the bottom
	three panels show the posterior distribution when the luminosity
	limit of the survey is $L > 2 \times 10^{44}\ [{\rm erg\
	    s^{-1}}]$. The vertical lines mark the true values of the
	parameters, $N = 1000, \alpha = 0,$ and $L^* = 10^{44}\ [{\rm erg\
	    s^{-1}}]$. The uncertainty on the parameters increases
	considerably when $L_{min} > L^*$, reflecting the fact that the
	bright end of the Schechter LF contains little information on
	$\alpha$ or $N$. \label{f-schechpost}}.
    \end{center}
  \end{figure}

  We used the Metropolis-Hastings algorithm described in
  \S~\ref{s-mha_schechter} to obtain a random draw of $\alpha,L^*$,
  and $N$ from the posterior probability distribution. The result was
  a set of $10^5$ random draws from the posterior probability
  distribution of $\alpha, L^*,$ and $N$. In Figure \ref{f-schechpost}
  we show the estimated posterior distribution of $\alpha, L^*$, and
  $N$ for both detection limits. While $L^*$ is fairly well
  constrained for both detection limits, the uncertainties on $\alpha$
  and $N$ are highly sensitive to whether the detection limit lies
  above or below $L^*$. In addition, the uncertainties on these
  parameters are not Gaussian, as is often assumed for the MLE.

\begin{figure}
  \begin{center}
    \includegraphics[scale=0.33,angle=90]{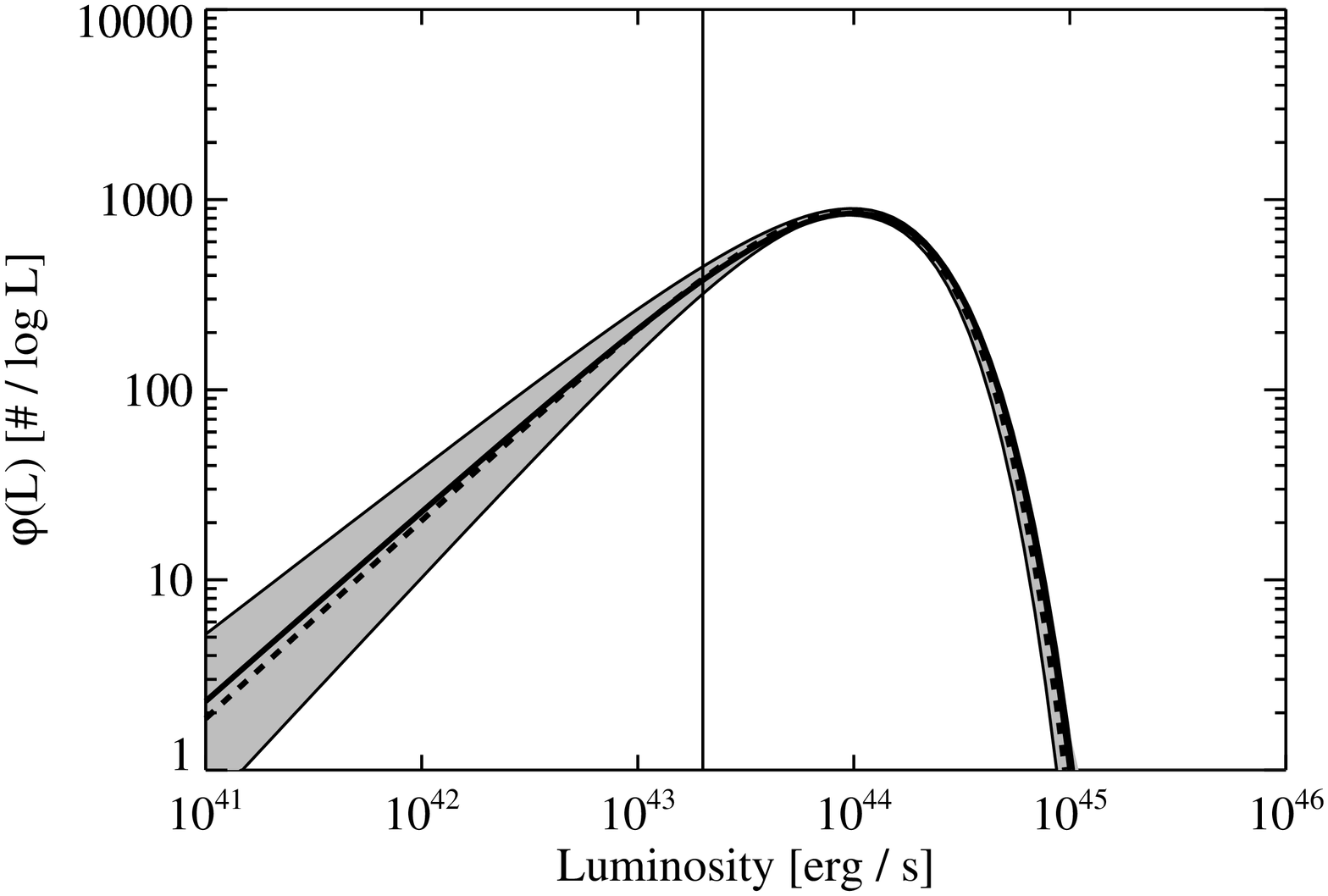}
    \includegraphics[scale=0.33,angle=90]{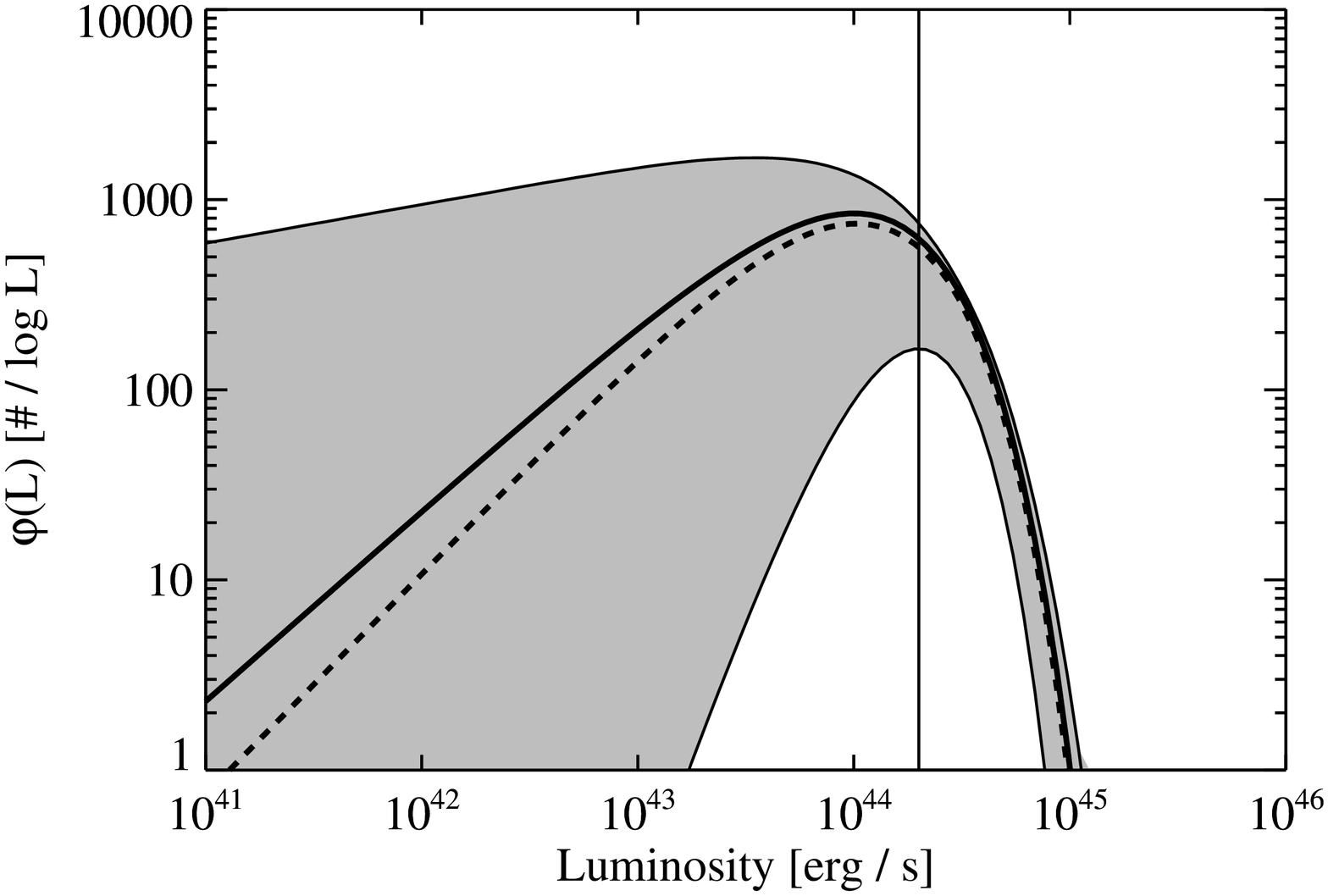}
    \caption{True value of the Schechter luminosity function (dashed
    line), compared with the best fit luminosity function calculated
    from the posterior median of the Schechter function parameters,
    $N, \alpha,$ and $L^*$ (solid line), from the simulated sample
    described in \S~\ref{s-schechter}. The left panel summarizes the
    posterior probability distributin of the LF when the luminosity
    limit is $L > 2 \times 10^{43}\ [{\rm erg\ s^{-1}}]$, and the
    right panel summarizes the posterior distribution of the LF when
    the luminosity limit is $L > 2 \times 10^{44}\ [{\rm erg\
    s^{-1}}]$. In both panels the shaded region contains $90\%$ of the
    posterior probability, and the vertical line marks the lower
    luminosity limit of the simulated survey. The uncertainty on the
    LF below the luminosity limit increases considerably when $L_{min}
    > L^*$, reflecting the fact that the bright end of the Schechter
    LF contains little information on $\alpha$ or $N$, and therefore
    contains little information on the faint end of the LF.
    \label{f-schechbounds}}.
  \end{center}
\end{figure}

  The random draws of $\alpha, L^*$, and $N$ can also be used to place
  constraints on the LF. This is done by computing Equation
  (\ref{eq-schechter}) for each of the random draws of $\alpha, L^*,$
  and $N$, and plotting the regions that contain, say, $90\%$ of the
  probability. In Figure \ref{f-schechbounds} we show the posterior
  median estimate of the LF, as well as the region containing 90\% of
  the posterior probability. As can be seen, the 90\% bounds contain
  the true value of the LF, and increase or decrease to reflect the
  amount of data available as a function of $L$. Furthermore, unlike
  the traditional MLE, these bounds do not rely on an assumption of
  Gaussian uncertainties, and therefore the confidence regions are
  valid for any sample size.

  \subsection{Comparison with Maximum-likelihood: Schechter Function}

  \label{s-schechter2}

  We also use Monte Carlo simulation to compare the Bayesian approach
  to maximum-likelihood for both the binomial and Poisson likelihood
  functions. We simulated 20 data sets for four types of surveys: (1)
  A large area shallow survey, (2) a large area medium depth survey,
  (3) a small area deep survey, and (4) a large area deep survey for
  rare objects, such as $z \sim 6$ quasars \citep[e.g.,][]{fan06}. For
  all four survey types we simulated quasars from a Schechter
  luminosity function with parameters the same as in
  \S~\ref{s-schechter}. For the large area shallow survey we used a
  total number of sources of $N = 10^5$, an area of $\Omega = 10^4\
  {\rm deg}^2$, and a lower luminosity limit of $L_{min} = 5 \times
  10^{44}\ {\rm erg\ s^{-1}}$. Only $n \sim 160$ sources are expected
  to be detected by this survey. For the large area medium depth
  survey we also used a LF normalization of $N = 10^5$ and area of
  $\Omega = 10^4\ {\rm deg}^2$, but instead used a lower luminosity
  limit of $L_{min} = 5 \times 10^{43}\ {\rm erg\ s^{-1}}$. The large
  area medium depth survey is expected to detect $n \sim 1.5 \times
  10^4$ sources. For the small area deep survey we used a survey area
  of $\Omega = 448\ {\rm arcmin}^2$, a LF normalization of $N = 5
  \times 10^7$ sources, and a lower luminosity limit of $L_{min} =
  10^{43}\ {\rm erg\ s^{-1}}$. This survey is expected to detect $n
  \sim 140$ sources. Finally, for the large area deep rare object
  survey we used an area of $\Omega = 10^4\ {\rm deg^2}$, a LF
  normalization of $N = 75$ sources, and a lower luminosity limit of
  $L_{min} = 10^{43}\ {\rm erg\ s^{-1}}$. Only $n \sim 16$ sources are
  expected to be detected by the rare object survey.

  We fit each of the 20 simulated data sets by maximum-likelihood for
  both the binomial and Poisson likelihood functions. The 95\%
  confidence intervals on the best-fit parameters were determined
  using 2000 bootstrap samples. We generate each bootstrap sample by
  randomly drawing n data points with replacement from the original
  sample, and then calculate the maximum-likelihood estimates for each
  bootstrap sample. We use the bootstrap to estimate the confidence
  intervals because the bootstrap does not assume that the errors are
  Gaussian, and because it is a common technique used in the LF
  literature. We estimate the 95\% confidence intervals directly from
  the 0.025 and 0.975 percentiles of the bootstrap sample. While
  bootstrap confidence intervals derived in this manner are known to
  be biased \citep[e.g.,][]{efron87,dav97}, additional corrections to
  the bootstrap samples are complicated. In addition, it is common
  practice to estimate bootstrap confidence intervals in this manner,
  and it is worth testing their accuracy. For the Bayesian approach,
  we used the MHA algorithm described in \S~\ref{s-mha_schechter} to
  simulate $5 \times 10^4$ random draws from the posterior
  distribution. The MHA algorithm was faster than fitting the 2000
  bootstrap samples using maximum likelihood.

  For each of the simulated samples, we counted the number of times
  that the true values of $N, \alpha,$ and $L^*$ were contained within
  the estimated 95\% confidence interval. The results are summarized
  in Table \ref{t-mlesim}. Because we estimated values of three
  parameters for 20 simulated data sets of 4 different types of
  surveys, we had 240 `trials' with probability of `success' $p =
  0.95$. If the estimated 95\% confidence regions corresponded to the
  true regions, then with $\approx 99\%$ probability between 220 and
  236 of the `trials' would fall within the estimated confidence
  region. For the binomial likelihood, the true value of a parameter
  was within the estimated 95\% confidence region only 210 times
  (88\%), and for the Poisson likelihood the true value of a parameter
  was within the estimated 95\% confidence region only 202 times
  (84\%). In contrast, the Bayesian approach was able to correctly
  constrain the true value of a parameter to be within the 95\%
  confidence region 233 times (97\%). Therefore, for our simulations
  confidence regions derived from bootstrapping the maximum-likelihood
  estimate are too narrow, while the confidence regions derived from
  the Bayesian method are correct.

  \begin{deluxetable}{lcccc}
    \tabletypesize{\scriptsize}
    \tablecaption{Schechter Function Confidence Intervals: Maximum-Likelihood vs. 
      Bayesian Inference\label{t-mlesim}}
    \tablewidth{0pt}
    \tablehead{ & Large Area, Shallow & Large Area, Medium & Small Area, Deep & Rare Object
    }
    \startdata
    \cutinhead{Normalization, $N$}
    Poisson  & 19 & 16 & 10 & 11 \\
    Binomial & 19 & 19 & 12 & 10 \\
    Bayesian & 20 & 19 & 19 & 20 \\
    \cutinhead{Faint End Power-law Slope, $k$}
    Poisson  & 18 & 17 & 18 & 17 \\
    Binomial & 18 & 19 & 18 & 18 \\
    Bayesian & 20 & 20 & 18 & 19 \\
    \cutinhead{Bright End Exponential Cut-off, $L^*$}
    Poisson  & 20 & 19 & 19 & 18 \\
    Binomial & 20 & 20 & 19 & 18 \\
    Bayesian & 20 & 20 & 18 & 20
    \enddata
    \tablecomments{Table \ref{t-mlesim} gives the number of times the true
      value of each parameter was contained within the estimated 95\%
      confidence interval for the simulated data sets described in
      \S~\ref{s-schechter2}. The results are reported seperately for each
      type of survey and Schechter function parameter. We simulated 20 data
      sets for each type of survey.}
  \end{deluxetable}

  Most of the failure in the maximum-likelihood confidence intervals
  came from the difficulty of the maximum-likelihood approach in
  constraining the LF normalization, $N$, for the small area deep
  survey and for the rare object survey. In particular, for these two
  surveys the bootstrap 95\% confidence intervals for both the
  binomial and Poisson likelihood function only contained the true
  value of $N$ roughly 50\% of the time. In general, there wasn't a
  significant difference among the simulated surveys in the ability of
  the three different statistical approaches to constrain $k$ and
  $L^*$ at 95\% confidence. However, the Poisson and binomial
  likelihood functions gave slightly different results for the larger
  area medium depth survey. For this survey the 95\% confidence
  intervals for the maximum-likelihood estimate derived from the
  Poisson distribution were somewhat smaller than those for the
  binomial distribution, only correcting including the true values of
  $N$, $k$, and $L^*$ roughly 85\% of the time. This is expected,
  because the Poisson distribution is the limit of the binomial
  distribution as the probability of including a source approaches
  zero; however, the detection probability for the large area medium
  depth survey is $\approx 0.15$.

  The results of our simulations imply that the Poisson likelihood
  function may lead to biased estimates of confidence intervals on the
  luminosity function parameters, so long as an astronomical survey is
  able to detect a significant fraction of the objects of
  interest. Use of the Poisson likelihood function may thus be
  problematic for the large astronomical surveys becoming common, so
  long as they are deep enough to achieve a moderate detection
  fraction. For the simulations performed here, using the Poisson
  likelihood function resulted in confidence intervals that are too
  narrow. However, these simulations were for a Schechter luminosity
  function, and other parameterizations may be affected
  differently. Considering that there are no obvious computational
  advantages to using the Poisson likelihood function, we recommend
  that the correct binomial likelihood function be used, as it is the
  correct form.

  \section{MIXTURE OF GAUSSIAN FUNCTIONS MODEL FOR THE LUMINOSITY FUNCTION}

  \label{s-smodel}

  In this section we describe a mixture of Gaussian functions model
  for the luminosity function. The mixture of Gaussians model is a
  common and well studied `non-parametric' model that allows
  flexibility when estimating a distribution, and is often employed
  when there is uncertainty regarding the specific functional form of
  the distribution of interest. The basic idea is that one can use a
  suitably large enough number of Gaussian functions to accurately
  approximate the true LF, even though the individual Gaussians have
  no physical meaning. In this sense, the Gaussian functions serve as
  a basis set of the luminosity function. As a result, we avoid the
  assumption of a more restrictive parametric form, such as a
  power-law, which can introduce considerable bias when extrapolating
  beyond the bounds of the observable data. We have not experimented
  with other luminosity function basis sets, although mixture models
  are very flexible and need not be limited to Gaussian functions.

  In this work we assume the mixture of Gaussian functions for the
  joint distribution of $\log L$ and $\log z$, as the logarithm of a
  strictly positive variable tends to more closely follow a normal
  distribution than does the untransformed variable. Therefore, we
  expect that a fewer number of Gaussians will be needed to accurately
  approximate the true LF, thus reducing the number of free
  parameters. Assuming a mixture of Gaussian functions for the joint
  distribution of $\log L$ and $\log z$ is equivalent to assuming a
  mixture of log-normal distributions for the distribution of $L$ and
  $z$. The mixture of $K$ Gaussian functions model for the $i^{\rm
  th}$ data point is
  \begin{equation}
    p(\log L_i, \log z_i|\pi, \mu, \Sigma) = \sum_{k=1}^{K} \frac{\pi_k}{2 \pi
      |\Sigma_k|^{1/2}} \exp \left[ -\frac{1}{2} ({\bf x}_i - \mu_k)^T
      \Sigma_k^{-1} ({\bf x}_i - \mu_k) \right],\ \ \theta = (\pi, \mu, \Sigma),
    \label{eq-mixmod}
  \end{equation}
  where $\sum_{k=1}^{K} \pi_k = 1$. Here, ${\bf x}_i = (\log L_i,\log
  z_i)$, $\mu_k$ is the 2-element mean (i.e., position) vector for the
  $k^{\rm th}$ Gaussian, $\Sigma_k$ is the $2 \times 2$ covariance
  matrix for the $k^{\rm th}$ Gaussian, and ${\bf x}^T$ denotes the
  transpose of ${\bf x}$. In addition, we denote $\pi = (\pi_1,
  \ldots, \pi_K), \mu = (\mu_1, \ldots, \mu_K)$, and $\Sigma =
  (\Sigma_1, \ldots, \Sigma_K)$. The variance in $\log L$ for Gaussian
  $k$ is $\sigma^2_{l,k} = \Sigma_{11,k}$, the variance in $\log z$
  for Gaussian $k$ is $\sigma^2_{z,k} = \Sigma_{22,k}$, and the
  covariance between $\log L$ and $\log z$ for Gaussian $k$ is
  $\sigma_{lz,k} = \Sigma_{12,k}$. In this work we consider the number
  of Gaussian components, $K$, to be specified by the researcher and
  fixed. If the number of Gaussian functions is also considered to be
  a free parameter, then methods exist for performing Bayesian
  inference on $K$ as well \citep[e.g.,][]{rich97}. Statistical
  inference on the luminosity functions studied in this work were not
  sensitive to the choice of $K$ so long as $K \ge 3$, and we conclude
  that values of $K \ge 3$ should be sufficient for most smooth and
  unimodal luminosity functions.

  Under the mixture model, the LF can be calculated from Equations
  (\ref{eq-phiconvert}) and (\ref{eq-mixmod}). Noting that $p(L,z) =
  p(\log L, \log z) / (L z (\ln 10)^2)$, the mixture of Gaussian
  functions model for the LF is
  \begin{equation}
    \phi(L,z|\theta,N) = \frac{N}{L z (\ln 10)^2} \left( \frac{dV}{dz} \right)^{-1}
    \sum_{k=1}^{K} \frac{\pi_k}{2 \pi |\Sigma_k|^{1/2}} \exp 
    \left[ -\frac{1}{2} ({\bf x} - \mu_k)^T \Sigma_k^{-1} ({\bf x} - \mu_k) \right] 
    \label{eq-mixlf},
  \end{equation}
  where, as before, ${\bf x} = (\log L, \log z)$. A mixture of
  Gaussian functions models was also used by \citet{blanton03} to
  estimate the $z = 0.1$ galaxy LF from the Sloan Digital Sky Survey
  (SDSS). Our mixture of Gaussian functions model differs from that
  used by \citet{blanton03} in that we do not fix the Gaussian
  function centroids to lie on a grid of values, and their individual
  widths are allowed to vary. This flexibility enables us to use a
  smaller number of Gaussian functions (typically $\sim 3-6$) to
  accurately fit the LF.

  \subsection{Prior Distribution}

  \label{s-prior}

  In this section we describe the prior distribution that we adopt on
  the mixture of Gaussian functions parameters. While one may be
  tempted to assumed a uniform prior on $\pi, \mu,$ and $\Sigma$, this
  will lead to an improper posterior, i.e., the posterior probability
  density does not integrate to one \citep{roeder97}. Therefore, a
  uniform prior cannot be used, and we need to develop a more
  informative prior distribution. Following \citet{roeder97}, we
  assume a uniform prior on $\pi_1,\ldots,\pi_K$ under the constraint
  that $\sum_{k=1}^K \pi_k = 1$; formally, this is a ${\rm
  Dirichlet}(1,\ldots,1)$ prior, where ${\rm
  Dirichlet}(\alpha_1,\ldots,\alpha_K)$ denotes a Dirichlet density
  with parameters $\alpha_1,\ldots,\alpha_K$. We give further details
  on the Dirichlet probability distribution in Appendix
  \S~\ref{a-densities}.
  
  Although our prior knowledge of the LF is limited, it is reasonable
  to assume a priori that the LF should be unimodal, i.e., that the LF
  should not exhibit multiple peaks. Currently, the observed
  distributions of galaxies and AGN are consistent with this
  assumption. For galaxies, unimodal luminosity function is implied by
  simple models of galaxy formation \citep[e.g.,][]{press74,schech76},
  and the luminosity functions for galaxies classified by morphology
  are not sufficiently separated to create multimodality
  \citep[e.g.,][]{nak03,scar07}. However, the luminosity function for
  AGN may be bimodal due to the possible existence of two difference
  accretion modes. This is largely due to the fact that the
  distribution of accretion rates relative to the Eddington rate,
  $\dot{m}$, is likely bimodal \citep{ho02,march04,hop06b,cao07} as a
  result of a transition from a radiatively inefficient flow to an
  efficient one at $\dot{m} \sim 0.01$
  \citep[e.g.,][]{jest05}. Because $L \propto M_{BH} \dot{m}$, the
  distribution of luminosities for AGN may be bimodal. If a survey is
  unable to detect those AGN in the faint radiatively inefficient
  mode, then the assumption of unimodality is violated and the
  estimated AGN luminosity function only refers to those AGN in the
  bright radiatively efficient mode.

\begin{figure}
  \begin{center}
    \includegraphics[scale=0.33,angle=90]{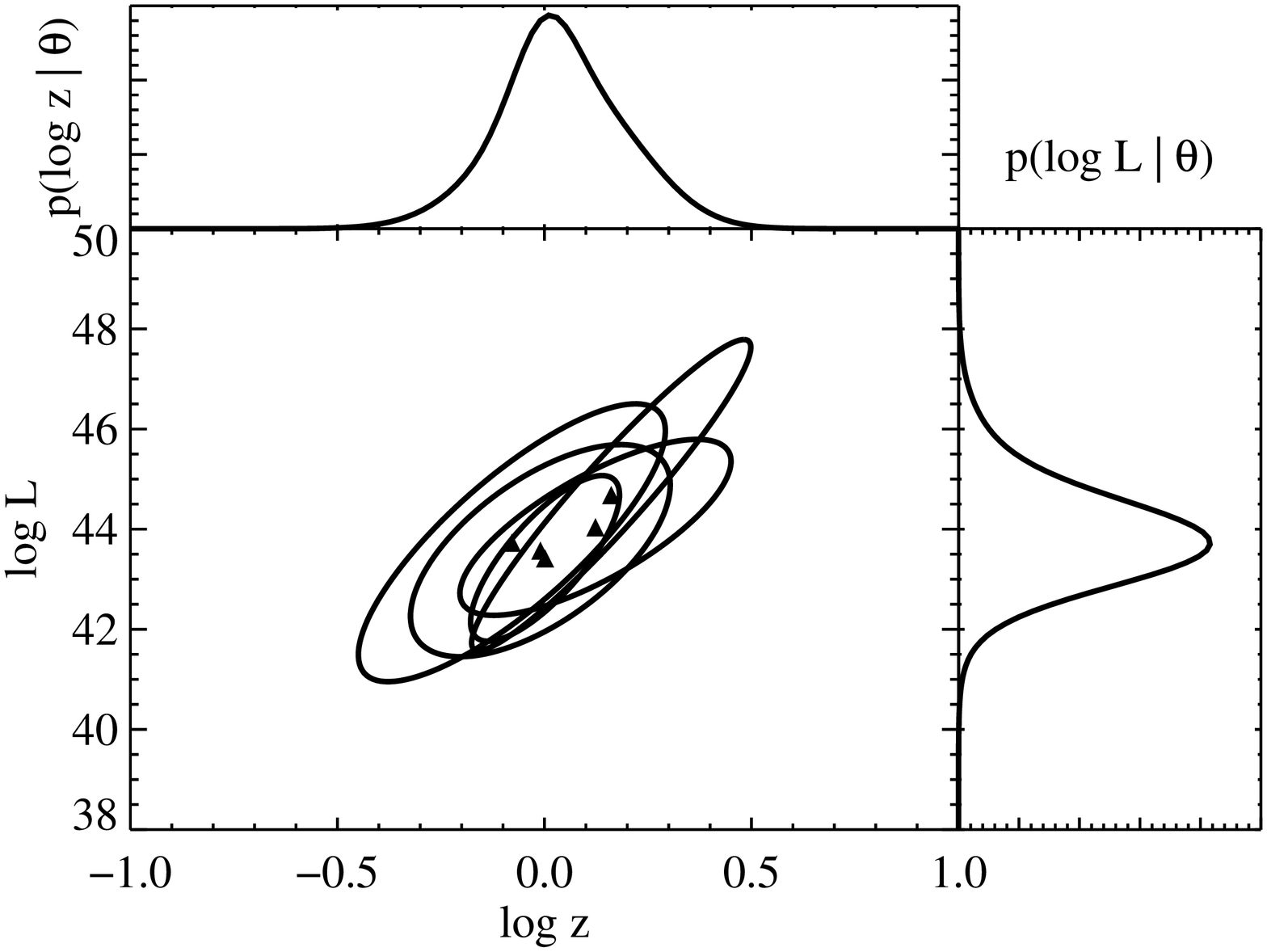}
    \includegraphics[scale=0.33,angle=90]{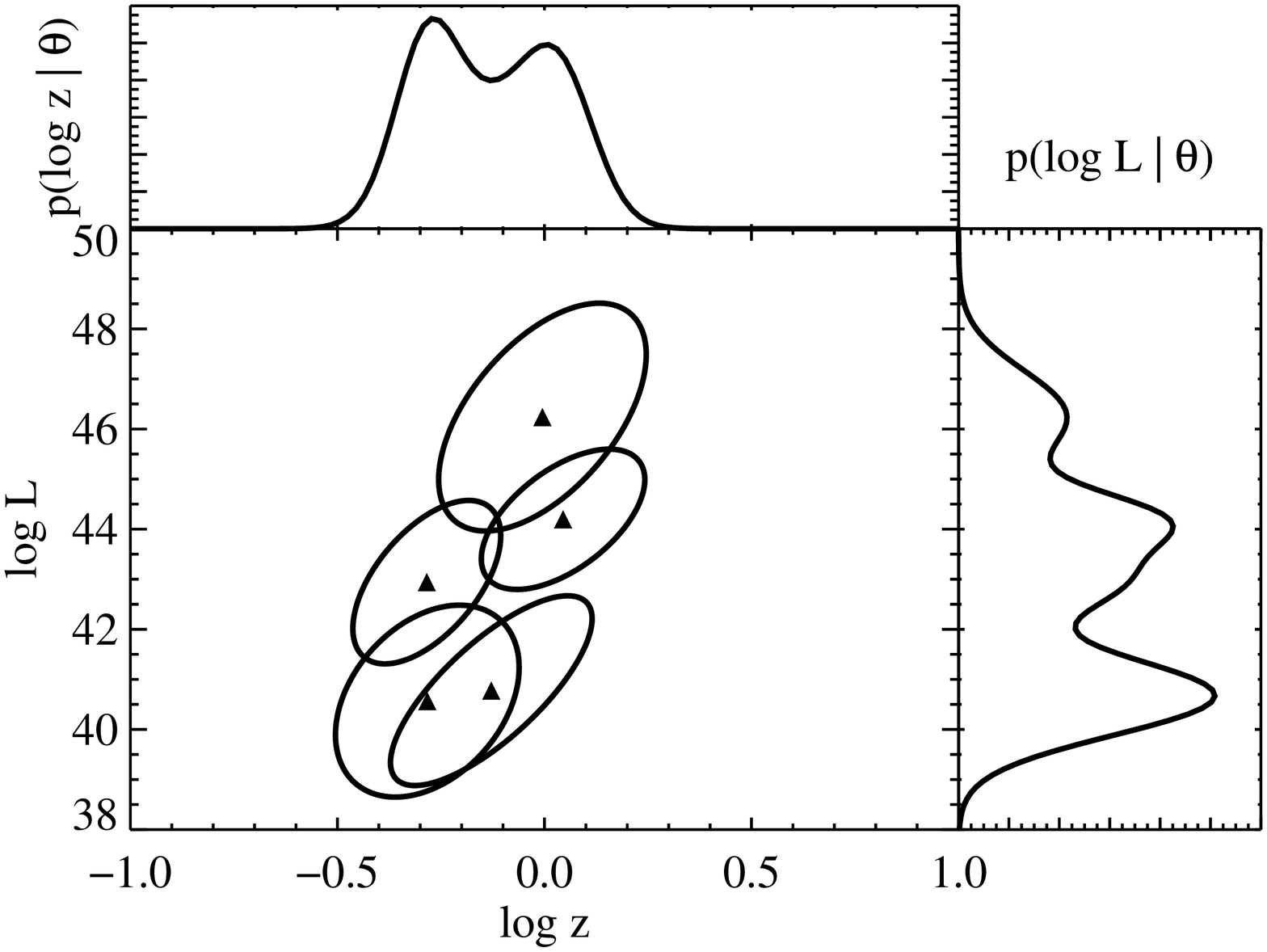}
    \caption{An illustration of our prior distribution for the
    Gaussian function parameters, for $K = 5$ Gaussian
    functions. Shown are a case when the Gaussian functions used in
    modelling the luminosity function are close together with respect
    to their covariances (left), and when the Gaussian functions are
    far apart with respect to their covariance matrices (right). The
    marginal distributions of $\log z$ are shown above the plots, and
    the marginal distributions of $\log L$ are shown to the right of
    the plots. When the Gaussian functions are close, the LF is
    unimodal, but when the Gaussian functions are far apart, the LF is
    multimodal. Because our prior distribution is constructed to place
    more probability on situations when the Gaussian functions are
    closer together with respect to their individual covariance
    matrices, it would place more probability on the situation shown
    in the left plot \emph{a priori}. Our prior therefore reflects our
    expectation that the LF should not exhibit multiple peaks (modes).
    \label{f-prior}}.
  \end{center}
\end{figure}

  To quantify our prior assumption that the LF is more likely to be
  unimodal, we construct our prior distribution to place more
  probability on situations where the individual Gaussian functions
  are close together in terms of their widths. In addition, we only
  specify the parametric form of the prior distribution, but allow
  the parameters of the prior distribution to vary and to be
  determined by the data. This allows our prior distribution to be
  flexible enough to have a minimal effect on the final results beyond
  conveying our prior consideration that the LF should be unimodal. We
  introduce our prior to place more probability on unimodal luminosity
  functions, to ensure that the posterior integrates to one, and to
  aid in convergence of the MCMC. Figure \ref{f-prior} illustrates the
  general idea that we are attempting to incorporate into our prior
  distribution. In this figure, we show a situation where the Gaussian
  functions are close together with respect to their widths, and far
  apart with respect to their widths. When the distances between the
  individual Gaussian functions, normalized by their covariance
  matrices (the measure of their `width'), is small, the LF is
  unimodal; however, when the distances between the Gaussian functions
  are large with respect to to their covariance matrices, the LF
  exhibits multiple modes. We construct a prior distribution that
  places less probability on the latter situation.

  Our prior on the Gaussian mean vectors and covariance matrices is
  similar to the prior described by \citet{roeder97}, but generalized
  to 2-dimensions.  For our prior, we assume an independent
  multivariate Cauchy distribution for each of the Gaussian means,
  $\mu$, with 2-dimensional mean vector $\mu_0$ and $2 \times 2$ scale
  matrix $T$. A Cauchy distribution is equivalent to a student's $t$
  distributions with 1 degree of freedom, and when used as a function
  in astronomy and physics it is commonly referred to as a Lorentzian;
  we describe the Cauchy distribution further in \S~\ref{a-densities}
  of the appendix. The scale matrix is chosen to be the harmonic mean
  of the Gaussian function covariance matrices:
  \begin{equation}
    T = \left( \frac{1}{K} \sum_{k=1}^K \Sigma_k^{-1} \right)^{-1}.
    \label{eq-priorscale}
  \end{equation}
  Qualitatively, this prior means that we consider it more likely that
  the centroids of the individual Gaussian functions should scatter
  about some mean vector $\mu_0$, where the width of this scatter
  should be comparable to the typical width of the individual Gaussian
  functions. The prior mean, $\mu_0$, is left unspecified and is an
  additional free parameter to be estimated from the data. We choose a
  Cauchy prior because the Cauchy distribution is heavy tailed, and
  therefore does not heavily penalize the Gaussian functions for being
  too far apart. As a result, the Cauchy prior is considered to be
  robust compared to other choices, such as the multivariate normal
  distribution.

  Because we use a random walk computational technique to explore the
  parameter space and estimate the posterior distribution, we find it
  advantageous to impose additional constraints on the Gaussian
  centroids. Both $\mu$ and $\mu_0$ are constrained to the region
  $\log L_{low} \leq \mu_{l,k} \leq \log L_{high}$ and $\log z_{low}
  \leq \mu_{z,k} \leq \log z_{high}$, where $\mu_{l,k}$ is the mean in
  $\log L$ for the $k^{\rm th}$ Gaussian, $\mu_{z,k}$ is the mean in
  $\log z$ for the $k^{\rm th}$ Gaussian. These constraints are
  imposed to keep the Markov chains (see \S~\ref{s-mha}) from
  `wandering' into unreasonable regions of the parameter space. The
  flux limit sets a lower limit on the luminosity of detected sources
  as a function of $z$, and therefore there is nothing in the observed
  data to `tell' the random walk that certain values of $\mu_l$ are
  unreasonable. For example, suppose our survey is only able to detect
  quasars with $L \gtrsim 10^{10} L_{\odot}$. Because of this, there
  is nothing in our data, as conveyed through the likelihood function,
  that says values of, say, $L \sim 10 L_{\odot}$ are unreasonable,
  and thus the Markov chains can get stuck wandering around values of
  $\mu_l \sim 1$. However, we know \emph{a priori} that values of
  $\mu_l \sim 1$ are unphysical, and therefore it is important to
  incorporate this prior knowledge into the posterior, as it is not
  reflected in the likelihood function. The values of these limits
  should be chosen to be physically reasonable. As an example, for the
  SDSS DR3 quasar LF with luminosities measured at $\lambda
  L_{\lambda} (2500$\AA$)$, it might be reasonable to take $L_{low} =
  10^{40}\ {\rm erg\ s^{-1}}$, $L_{high} = 10^{48}\ {\rm erg\
  s^{-1}}$, $z_{low} = 10^{-4}$, and $z_{high} = 7$.

  Generalizing the prior of \citet{roeder97}, we assume independent
  inverse Wishart priors on the individual Gaussian covariance
  matrices with $\nu = 1$ degrees of freedom, and common scale matrix
  $A$. We give a description of the Wishart and inverse Wishart
  distributions in \S~\ref{a-densities}. This prior states that the
  individual $\Sigma_k$ are more likely to be similar rather than
  different. The common scale matrix, $A$, is left unspecified so it
  can adapt to the data. As with $\mu$, we recommend placing upper and
  lower limits on the allowable values of dispersion in $\log L$ and
  $\log z$ for each Gaussian..

  Mathematically, our prior is
  \begin{eqnarray}
    p(\pi,\mu,\Sigma,\mu_0,A) & \propto & \prod_{k=1}^K
    p(\mu_k|\mu_0,\Sigma) p(\Sigma_k|A) \\ 
    & \propto & \prod_{k=1}^K {\rm Cauchy}_2(\mu_k|\mu_0,T) \mbox{\rm Inv-Wishart}_1(\Sigma_k|A),
    \label{eq-prior}
  \end{eqnarray}
  under the constraints given above. Here, ${\rm Cauchy}_2
  (\mu_k|\mu_0,T)$ denotes a 2-dimensional Cauchy distribution as a
  function of $\mu_k$, with mean vector $\mu_0$ and scale matrix
  $T$. In addition, Inv-Wishart$_1 (\Sigma_k|A)$ denotes an inverse
  Wishart density as a function of $\Sigma_k$, with one degree of
  freedom and scale matrix $A$. We have also experimented with using a
  uniform prior on the parameters, constricted to some range. In
  general, this did not change our constraints on the LF above the
  flux limit, but resulted in somewhat wider confidence regions on the
  LF below the flux limit. This is to be expected, since our adopted
  Cauchy prior tends to restrict the inferred LF to be unimodal, and
  therefore limits the number of possible luminosity functions that
  are considered to be consistent with the data.

  \subsection{Posterior Distribution for Mixture of Gaussians Model}

  Now that we have formulated the prior distribution, we can calculate
  the posterior distribution for the mixture of Gaussians model of
  $\phi(L,z)$. Because we have formulated the mixture model for the LF
  in terms of $\log L$ and $\log z$, the marginal posterior
  distribution of $\theta$ is
  \begin{equation}
    p(\theta,\mu_0,A|\log L_{obs}, \log z_{obs}) \propto p(\theta,\mu_0,A) 
    \left[p(I=1|\theta) \right]^{-n} \prod_{i \in {\cal A}_{obs}} 
    p(\log L_i,\log z_i|\theta),\ \ \theta = (\pi,\mu,\Sigma), \label{eq-thetapost_log}
  \end{equation}
  where $p(\theta,\mu_0,A)$ is given by Equation (\ref{eq-prior}),
  $p(\log L_i,\log z_i|\theta)$ is given by Equation
  (\ref{eq-mixmod}), and
  \begin{equation}
    p(I=1|\theta) = \int_{-\infty}^{\infty} \int_{-\infty}^{\infty} 
    p(I = 1|\log L,\log z) p(\log L,\log z|\theta)\ d\log L\ d\log z
    \label{eq-selprob_log}
  \end{equation}
  is the probability of including a source, given the model parameters
  $\theta$. The conditional posterior distribution of $N$ given $\pi,
  \mu,$ and $\Sigma$ is given by inserting Equation
  (\ref{eq-selprob_log}) into (\ref{eq-npost}). The complete joint
  posterior distribution is then
  \begin{equation}
    p(\theta,N,\mu_0,A|\log L_{obs},\log z_{obs}) \propto 
    p(N|\theta,n) p(\theta,\mu_0,A|\log L_{obs},\log z_{obs}).
    \label{eq-mixmod_post}
  \end{equation}
 
  \section{USING MARKOV CHAIN MONTE CARLO TO ESTIMATE THE POSTERIOR 
    DISTRIBUTION OF THE LUMINOSITY FUNCTION}

  \label{s-mha}
  
  For our statistical model, $\mu_0$ has 2 free parameters, $A$ has 3
  free parameters, and each of the $K$ Gaussian components has 6 free
  parameters. Because the values of $\pi$ are constrained to sum to
  one, there are only $6K - 1$ free parameters for the Gaussian
  mixture model. The number of free parameters in our statistical
  model is therefore $6K + 4$. The large number of parameters
  precludes calculation of the posterior on a grid of $\pi, \mu,
  \Sigma, \mu_0, A,$ and $N$. Furthermore, the multiple integrals
  needed for marginalizing the posterior, and thus summarizing it, are
  numerically intractable. Because of this, we employ Markov Chain
  Monte Carlo (MCMC) to obtain a set of random draws from the
  posterior distribution. A Markov chain is a random walk, where the
  probability distribution of the current location only depends on the
  previous location. To obtain random numbers generated from the
  posterior distribution, one constructs a Markov chain that performs
  a random walk through the parameter space, where the Markov chain is
  constructed to eventually converge to the posterior
  distribution. Once convergence is reached, the values of the Markov
  chain are saved at each iteration, and the values of these locations
  can be treated as a random draw from the posterior
  distribution. These draws may then be used to estimate the posterior
  distribution of $\phi(L,z)$, and thus an estimate of the LF and its
  uncertainty can be obtained.

  In this work we use the Metropolis-Hastings algorithm
  \citep[MHA,][]{metro49,metro53,hast70} to perform the MCMC. We
  describe the particular MHA we employ for Bayesian inference on the
  LF; however for a more general and complete description of the MHA,
  we refer the reader to \citet{chib95} or \citet{gelman04}. We use
  the MHA to obtain a set of random draws from the marginal posterior
  distribution of $\theta$, given by Equation
  (\ref{eq-thetapost_log}). Then, given these random draws of
  $\theta$, random draws for $N$ may be obtained directly from the
  negative binomial distribution (see Eq.[\ref{eq-npost}] and
  \S~\ref{a-densities} of the appendix).

\begin{figure}
  \begin{center}
    \scalebox{0.7}{\rotatebox{90}{\plotone{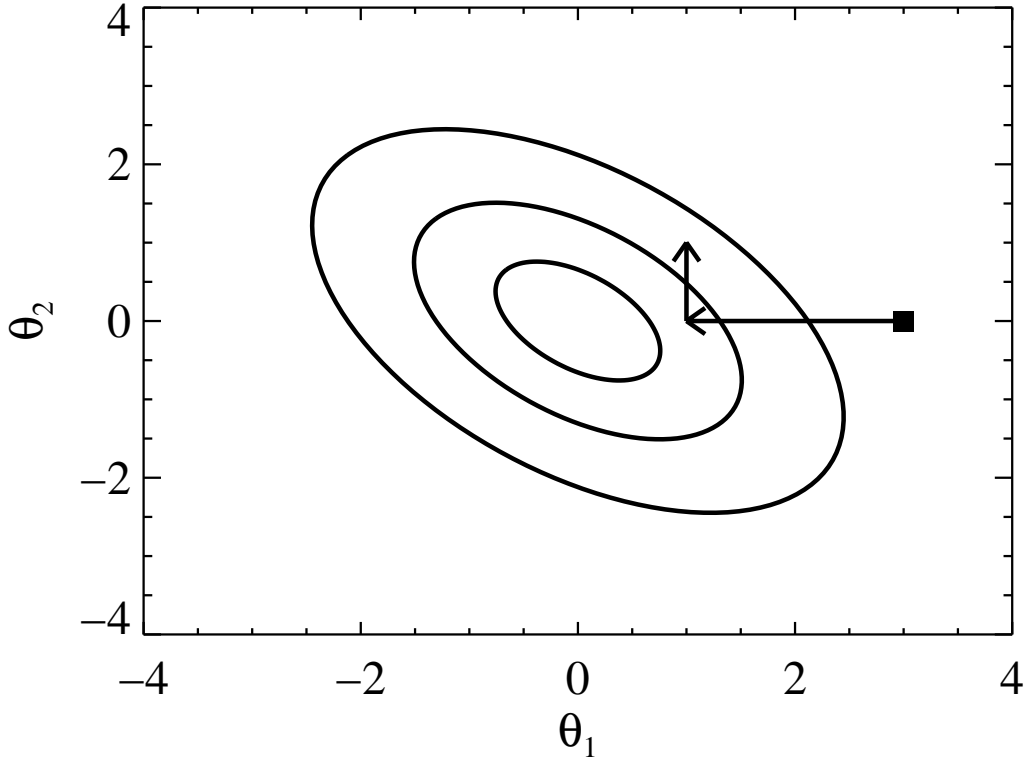}}}
    \caption{Schematic diagram illustrating the random walk
    Metropolis-Hastings algorithm. The posterior probability
    distribution is illustrated by the contours, and the random walk
    is initially at the position marked with a square. A new proposed
    value of $\theta_1$ is randomly drawn, marked by the arrow
    pointing to the left. Because the proposed value of $\theta_1$ is
    at a location with higher posterior probability, the new value of
    $\theta_1$ is saved, and the random walk `jumps' to the position
    marked by the arrow. Then, a new proposal for $\theta_2$ is
    randomly drawn, marked by the arrow pointing upward. Because this
    proposed value of $\theta_2$ is at a location with lower posterior
    probability, it is only accepted with probability equal to the
    ratio of the values of the posterior at the proposed position and
    the current position. If the proposed value is kept, then the new
    value of $\theta_2$ is saved, otherwise the current value of
    $\theta_2$ is saved. Next, a proposed value of $\theta_1$ is
    randomly drawn, and the process repeats, creating a random walk
    through the parameter space. Because the amount of time that the
    random walk spends in any given bin in $\theta_1$ and $\theta_2$
    is proportional to the posterior probability distribution, after
    the random walk has converged, the values of $\theta_1$ and
    $\theta_2$ from the random walk may be treated as a random draw
    from the posterior distribution.
 \label{f-mha}}
  \end{center}
\end{figure}

  The basic idea behind the MHA is illustrated in Figure \ref{f-mha}
  for the special case of a symmetric jumping distribution. First, one
  starts with an initial guess for $\theta$. Then, at each iteration a
  proposed value of $\theta$ is randomly drawn from some `jumping'
  distribution. For example, this jumping distribution could be a
  normal density with some fixed covariance matrix, centered at the
  current value of $\theta$. Then, if the proposed value of $\theta$
  improves the posterior, it is stored as the new value of
  $\theta$. Otherwise, it is stored as the new value with probability
  equal to the ratio of the values of the posterior distribution at
  the proposed and current value of $\theta$. If the proposed value of
  $\theta$ is rejected, then the value of $\theta$ does not change,
  and the current value of $\theta$ is stored as the `new' value of
  $\theta$. The process is repeated until convergence. If the jumping
  distribution is not symmetric, then a correction needs to be made to
  the acceptance rule in order to account for asymmetry in the jumping
  distribution. A jumping distribution is symmetric when the
  probability of jumping from a current value $\theta$ to a new value
  $\theta^*$ is the same as jumping from $\theta^*$ to $\theta$. For
  example, the normal distribution is symmetric, while the log-normal
  distribution is not.

  \subsection{Metropolis-Hastings Algorithm for Schechter Luminosity Function}

  \label{s-mha_schechter}

  Before describing our MHA algorithm for the mixture of Gaussian
  functions model, we describe a simpler MHA algorithm for the
  Schechter function model given by Equation (\ref{eq-schechter}) in
  \S~\ref{s-schechter}. We do this to illustrate the MHA using a more
  familiar luminosity function. An MHA for obtaining random draws of
  $\alpha, L^*,$ and $N$ from Equation (\ref{eq-schechpost}) is:
  \begin{enumerate}
  \item
    \label{i-sstart}
    Start with an initial value of $\alpha$ and $L^*$, denoted as
    $\tilde{\alpha}$ and $\tilde{L}^*$. A good initial value is the
    maximum-likelihood estimate.
  \item
    \label{i-slstar}
    Draw a proposal value of $\log L^*$ from a normal distribution
    centered on the current value of $\log L^*$, $\log
    \tilde{L}^*$. The variance in the jumping distribution of $\log
    L^*$ should be fixed at the beginning of the MHA. A larger jumping
    variance will lead to jumps that travel greater distances, but
    will then lead to lower MHA acceptance rates. The value of the
    jumping variance should be tuned to give acceptance rates $\sim
    0.4$.  We use a normal jumping distribution to vary $\log L^*$
    because $\log L^*$ is defined on $(-\infty,\infty)$, while $L^*$
    is only defined on $(0,\infty)$. While we could use a jumping
    distribution to directly vary $L^*$, it is not always easy to
    simulate random variables directly from distributions that are
    only defined for $L^* > 0$.

    Denoting the proposal value of $L^*$ as $\hat{L}^*$, calculate the
    ratio
    \begin{equation}
      r_{L^*} = \frac{\hat{L}^* p(\tilde{\alpha},\hat{L}^*|L_{obs})}
      {\tilde{L}^* p(\tilde{\alpha}, \tilde{L}^*|L_{obs})}
      \label{eq-rlstar}
    \end{equation}
    Here, $p(\alpha, L^*|L_{obs})$ is the posterior distribution for
    the Schechter function, given by Equation (\ref{eq-schechpost}).
    If $r_{L^*} > 1$, then keep the proposal and set $\tilde{L}^* =
    \hat{L}^*$. If $r_{L^*} < 1$, then draw a random number $u$
    uniformly distributed between 0 and 1. If $u < r_{L^*}$, then keep
    the proposal and set $\tilde{L}^* = \hat{L}^*$. Otherwise, the
    proposal is rejected and the value of $\tilde{L}^*$ is
    unchanged. The factor of $\hat{L}^* / \tilde{L}^*$ is necessary in
    Equation (\ref{eq-rlstar}) in order to correct for the asymmetry in
    the log-normal jumping distribution.
  \item
    \label{i-salpha}
    Draw a proposal value of $\log k = \log (\alpha + 1)$ from a
    normal distribution centered at the current value of $\log k$,
    $\log \tilde{k} = \log (\tilde{\alpha} + 1)$. Similar to the MHA
    step for $L^*$, we use a normal jumping distribution to vary $\log
    k$ because $\log k$ is defined on $(-\infty,\infty)$, while
    $\alpha$ is only defined on $(-1,\infty)$.

    Denoting the proposal value of $k$ as $\hat{k}$, the proposal
    value of $\hat{\alpha}$ is $\hat{\alpha} = \hat{k} - 1$. Using the
    values of $\tilde{\alpha}$ and $\hat{\alpha}$, calculate the ratio
    \begin{equation}
      r_{\alpha} = \frac{\hat{k} p(\hat{\alpha},\tilde{L}^*|L_{obs})}
      {\tilde{k} p(\tilde{\alpha}, \tilde{L}^*|L_{obs})}
      \label{eq-ralf}
    \end{equation}
    If $r_{\alpha} > 1$, then keep the proposal and set
    $\tilde{\alpha} = \hat{\alpha}$. If $r_{\alpha} < 1$, then draw a
    random number $u$ uniformly distributed between 0 and 1. If $u <
    r_{\alpha}$, then keep the proposal and set $\tilde{\alpha} =
    \hat{\alpha}$. Otherwise, the proposal is rejected and the value
    of $\tilde{\alpha}$ is unchanged. As with the MHA step for $L^*$,
    the factor of $\hat{k} / \tilde{k}$ is necessary in Equation
    (\ref{eq-ralf}) in order to correct for the asymmetry in the
    log-normal jumping distribution.
  \item
    \label{i-srepeat}
    Repeat steps (\ref{i-slstar})--(\ref{i-salpha}) until the MHA
    algorithm converges. Techniques for monitoring convergence are
    described in \citet{gelman04}. After convergence, use Equation
    (\ref{eq-npost}) to directly simulate random draws of the LF
    normalization, $N$, for each simulated value of $\alpha$ and $L^*$
    obtained from the above random walk. Equation (\ref{eq-npost}) has
    the form of a negative binomial distribution, and a method for
    simulated random variables from the negative binomial distribution
    is described in \S~\ref{a-densities} of the appendix.
  \end{enumerate}

  \subsection{Metropolis-Hastings Algorithm for the Mixture of Gaussian Functions
    Luminosity Function}

  \label{s-mha_mixture}

  Our MHA for the mixture of Gaussian functions model is a more
  complex version of that used for the Schechter function model. As
  before, we denote the current value of a parameter by placing a
  $\tilde{}$ over its symbol, and we denote the proposed value by
  placing a $\hat{}$ over its symbol. For example, if one were
  updating $\pi$, then $\tilde{\pi}$ denotes the current value of
  $\pi$ in the random walk, and $\hat{\pi}$ denotes the proposed value
  of $\pi$. We will only update one parameter at a time, so, if we are
  drawing a proposal for $\pi$, the current value of $\theta$ is
  denoted as $\tilde{\theta} = (\tilde{\pi}, \tilde{\mu},
  \tilde{\Sigma})$, and the proposed value of $\theta$ is denoted as
  $\hat{\theta} = (\hat{\pi}, \tilde{\mu}, \tilde{\Sigma})$.

  Our MHA for the mixture of Gaussian functions model is:
  \begin{enumerate}
  \item
    \label{i-mstart}
    Start with initial guesses for $\pi, \mu, \Sigma, \mu_0, A,$ and
    $T$.
  \item
    \label{i-pi}
    Draw a proposal value for $\pi$ from a ${\rm
    Dirichlet}(\tilde{g}_1,\ldots,\tilde{g}_K)$ density, where
    $\tilde{g}_k = c_{\pi} n \tilde{\pi}_k + 1$, $n$ is the number of
    sources in the survey, and $c_{\pi}$ is a fixed positive constant
    that controls how far the `jumps' in $\pi$ go. Because $c_{\pi}$
    controls the variance of the Dirichlet density, a smaller value of
    $c_{\pi}$ produces values of $\hat{\pi}$ that are further from
    $\tilde{\pi}$. The value of $c_{\pi}$ should be chosen so that
    about $15$--$40\%$ of the MHA proposals are accepted.

    After drawing a proposal for $\pi$, calculate the value of the
    posterior distribution at the new value of $\theta$,
    $\hat{\theta}$, and at the old value of $\theta$,
    $\tilde{\theta}$. Then, use these values to calculate the ratio
    \begin{equation}
      r_{\pi} = \frac{{\rm Dirichlet}(\tilde{\pi}|\hat{g})}{{\rm Dirichlet}(\hat{\pi}|\tilde{g})}
      \frac{p(\hat{\theta}|L_{obs},z_{obs})}{p(\tilde{\theta}|L_{obs},z_{obs})}, \label{eq-pijump}
    \end{equation}
    where $\hat{g} = c_{\pi} n \hat{\pi}_1 + 1,\ldots,c_{\pi} n
    \hat{\pi}_K + 1$. The ratio of Dirichlet densities in Equation
    (\ref{eq-pijump}) corrects the MHA acceptance rule for the
    asymmetry in the Dirichlet jumping distribution. If $r_{\pi} \geq
    1$ then keep the proposed value of $\pi$: $\tilde{\pi} =
    \hat{\pi}$. Otherwise keep the proposal with probability
    $r_{\pi}$. This is done by drawing a uniformly distributed random
    variable between 0 and 1, denoted by $u$. If $u < r_{\pi}$, then
    set $\tilde{\pi} = \hat{\pi}$. If $u > r_{\pi}$ then keep the
    current value of $\pi$.

    Methods for simulating from the Dirichlet distribution, as well as
    the functional form of the Dirichlet distribution, are given in
    \S~\ref{a-densities}.
  \item
    \label{i-mu}
    For each Gaussian function, draw a proposal for $\mu_k$ by drawing
    $\hat{\mu}_k \sim N_2(\tilde{\mu}_k,V_k)$, where $V_k$ is some set
    covariance matrix. Because the jumping density is symmetric, the
    MHA acceptance ratio is just given by the ratio of the posterior
    distributions at the proposed and current value of $\mu_k$:
    $r_{\mu} = p(\hat{\theta}|L_{obs},z_{obs}) /
    p(\tilde{\theta}|L_{obs},z_{obs})$. If $r_{\mu} \geq 1$ then set
    $\tilde{\mu}_k = \hat{\mu}_k$, otherwise set $\tilde{\mu}_k =
    \hat{\mu}_k$ with probability $r_{\mu}$. The MHA update should be
    performed separately for each Gaussian. The covariance matrix of
    the jumping kernel, $V_k$, should be chosen such that $\sim 30\%$
    of the MHA jumps are accepted.

    Since we have constructed the prior distribution with the
    constraint $\log L_{low} \leq \mu_{l,k} \leq \log L_{high}$ and
    $\log z_{min} \leq \mu_{z,k} \leq \log z_{high}$ for all $k$, any
    values of $\hat{\mu}_k$ that fall outside of this range should
    automatically be rejected.
  \item
    \label{i-covar}
    For each Gaussian, draw a proposal for $\Sigma_k$ by drawing
    $\hat{\Sigma}_k \sim$ Wishart$_{\nu_k}(\tilde{\Sigma}_k / \nu_k)$,
    where $\nu_k$ is some set degrees of freedom. Larger values of
    $\nu_k$ will produce values of $\hat{\Sigma}_k$ that are more
    similar to $\tilde{\Sigma}_k$. The MHA acceptance ratio is
    \begin{equation}
      r_{\Sigma} = \left(\frac{|\tilde{\Sigma}_k|}{|\hat{\Sigma}_k|}\right)^{\nu_k - 3/2}
      \exp\left\{-\frac{\nu_k}{2} tr\left[(\hat{\Sigma}_k)^{-1} \tilde{\Sigma}_k - 
	\tilde{\Sigma}_k^{-1} \hat{\Sigma}_k \right] \right\}
      \frac{p(\hat{\theta}|x_{obs})}{p(\tilde{\theta}|x_{obs})},
      \label{eq-sigjump}
    \end{equation}
    where $tr(\cdot)$ denotes the trace of a matrix. If $r_{\Sigma}
    \geq 1$ then set $\tilde{\Sigma}_k = \hat{\Sigma}_k$, otherwise set
    $\tilde{\Sigma}_k = \hat{\Sigma}_k$ with probability $r_{\Sigma}$. The
    MHA update should be performed separately for each Gaussian. The
    degrees of freedom of the jumping kernel, $\nu_k$, should be
    chosen such that $\sim 15$--$40\%$ of the MHA jumps are accepted.

    If there are any bounds on $\Sigma_k$ incorporated into the prior
    distribution, then values of $\Sigma_k$ that fall outside of this
    range should automatically be rejected. Methods for simulating
    from the Wishart distribution, as well as the functional form of
    the Wishart distribution, are given in \S~\ref{a-densities}.
  \item
    \label{i-mu0}
    Draw a proposal for the prior parameter $\mu_0$ as $\hat{\mu}_0
    \sim N_2(\tilde{\mu}_0,V_0)$. The acceptance ratio only depends on
    the prior distribution and is
    \begin{equation}
      r_0 = \left[ \prod_{k=1}^K \frac{{\rm Cauchy}_2(\mu_k|\hat{\mu}_0,T)}
	{{\rm Cauchy}_2(\mu_k|\tilde{\mu}_0,T)} \right]
      \left[ \frac{\int_{\log L_{low}}^{\log L_{high}} \int_{\log z_{low}}^{\log z_{high}} 
	  {\rm Cauchy}_2(\mu_k|\tilde{\mu}_0,T)\ d\mu_k}
	{\int_{\log L_{low}}^{\log L_{high}} \int_{\log z_{low}}^{\log z_{high}} 
	  {\rm Cauchy}_2(\mu_k|\hat{\mu}_0,T)\ d\mu_k} \right]^K. \label{eq-mu0jump}
    \end{equation}
    Here, $T$ is given by Equation (\ref{eq-priorscale}) and the
    integrals are needed because of the prior constraints on $\mu$. If
    $r_0 \geq 1$ then set $\tilde{\mu}_0 = \hat{\mu}_0$, otherwise set
    $\tilde{\mu}_0 = \hat{\mu}_0$ with probability $r_0$. We have
    found a good choice for $V_0$ to be the sample covariance matrix
    of $\tilde{\mu}$.
  \item
    \label{i-aupdate}
    Finally, update the value of $A$, the common scale matrix. Because
    we can approximately calculate the conditional distribution of
    $A$, given $\Sigma$, we can directly simulate from
    $p(A|\Sigma)$. Directly simulating from the conditional
    distributions is referred to as a Gibbs sampler. We perform a
    Gibbs update to draw a new value of $\tilde{A}$:
    \begin{eqnarray}
      \hat{A} & \sim & {\rm Wishart}_{\nu_A}(S) \label{eq-ajump} \\
      \nu_A & = & K + 3 \\
      S & = & \left( \sum_{k=1}^K \tilde{\Sigma}_k^{-1} \right)^{-1}.
    \end{eqnarray}
    For the Gibbs sampler update, we do not need to calculate an
    acceptance ratio, and every value of $\hat{A}$ is accepted: $\tilde{A}
    = \hat{A}$. If there are any prior bounds set on $\Sigma$, then this
    is technically only an approximate Gibbs update, as it ignores the
    constraint on $\Sigma$. A true MHA update would account for the
    constraint on $\Sigma$ by renormalizing the conditional
    distribution appropriately; however, this involves a triple
    integral that is expensive to compute. If there are prior bounds
    on $\Sigma$, then Equation (\ref{eq-ajump}) is approximately
    correct, and ignoring the normalization in $p(A|\Sigma)$ does not
    did not have any effect on our results.
  \end{enumerate}
  Steps \ref{i-pi}--\ref{i-aupdate} are repeated until the MCMC
  converges, where one saves the values of $\tilde{\theta}$ at each
  iteration. After convergence, the MCMC is stopped, and the values of
  $\tilde{\theta}$ may be treated as a random draw from the marginal
  posterior distribution of $\theta$, $p(\theta|\log L_{obs}, \log
  z_{obs})$.  Techniques for monitoring convergence of the Markov
  Chains are described in \citet{gelman04}. If one wishes to assume a
  uniform prior on $\mu$ and $\Sigma$, constrained within some set
  range, instead of the prior we suggest in \S~\ref{s-prior}, then
  only steps \ref{i-pi}--\ref{i-covar} need to be performed. Given the
  values of $\theta$ obtained from the MCMC, one can then draw values
  of $N$ from the negative binomial density
  (cf. Eq.[\ref{eq-npost}]). In \S~\ref{a-densities} we describe how
  to simulate random variables from a negative binomial
  distribution. The speed of our MHA algorithm depends on the sample
  size and the programming language. As a rough guide, on a modern
  computer our MHA can take a couple of hours to converge for sample
  sizes of $\sim 1000$, and our MHA can take as long as a day or two
  to converge for sample sizes $\sim 10^4$.

  When performing the MCMC it is necessary to perform a `burn-in'
  stage, after which the Markov chains have approximately converged to
  the posterior distribution. The values of $\theta$ from the MCMC
  during the burn-in stage are discarded, and thus only the values of
  $\theta$ obtained after the burn-in stage are used in the
  analysis. We have found it useful to perform $\sim 10^4$ iterations
  of burn-in, although this probably represents a conservative
  number. In addition, the parameters for the MHA jumping
  distributions should be tuned during the burn-in stage. In
  particular, the parameters $\Sigma_{\alpha}, \sigma^2_{\sigma_l},
  c_{\pi}, \Sigma_{\mu,k},$ and $\nu_k$ should be varied within the
  burn-in stage to make the MHA more efficient and have an acceptance
  rate of $\sim 0.15$--$0.4$ \citep{gelman95}. These jumping
  distribution parameters cannot be changed after the burn-in
  stage. \citet{jasra05} and \citet{neal96} described additional
  complications and considerations developing MHAs for mixture models.

  Some post processing of the Markov chains is necessary. This is
  because some chains can get `stuck' wandering in regions far below
  flux limit, likely in the presence of a local maximum in the
  posterior. While such chains will eventually converge and mix with
  the other chains, they do not always do so within the finite number
  of iterations used when running the random walk MHA. We argued in
  \S~\ref{s-prior} that the Gaussian centroids should be limited to
  some specified range in L and z to prevent the chains from getting
  stuck. However, this is only a partial fix, as the minimum
  luminosity for the Gaussian function means may be significantly
  fainter than the flux limit at a given redshift, $L_{lim}(z); i.e.,
  \mu_l \ge \log L_{low}$ and $L_{low} < L_{lim}(z)$. In general, we
  have found that divergent chains are easy to spot. Because the
  divergent chains usually get stuck in regions far below the flux
  limit, they correspond to luminosity functions with implied
  extremely low detection probabilities, i.e., $p(I=1|\theta) \ll
  1$. As a result, the random draws of $N$ from the posterior for
  these chains tend to have values that are too high and far removed
  from the rest of the posterior distribution of $N$. The divergent
  chains are therefore easily found and removed by inspecting a
  histogram of $\log N$. In fact, we have found that the divergent
  chains often become too large for the long integer format used in
  our computer routines, and therefore are returned as negative
  numbers. Because negative values of $N$ are unphysical, it is easy
  to simply remove such chains from the analysis.

  Having obtained random draws of $N$ and $\theta$ from
  $p(\theta,N|\log L_{obs},\log z_{obs})$, one can then use these
  values to calculate an estimate of $\phi(L,z)$, and its
  corresponding uncertainty. This is done by inserting the MCMC values
  of $\theta$ and $N$ directly into Equation (\ref{eq-mixlf}). The
  posterior distribution of $\phi(L,z)$ can be estimated for any value
  of $L$ and $z$ by plotting a histogram of the values of $\phi(L,z)$
  obtained from the MCMC values of $\theta$ and $N$. In
  \S~\ref{s-sim}, we illustrate in more detail how to use the MHA
  results to perform statistical inference on the LF.

  \section{APPLICATION TO SIMULATED DATA}

  \label{s-sim}

  As an illustration of the effectiveness of our method, we applied it
  to a simulated data set. We construct a simulated sample, and then
  recover the luminosity function based on our mixture of Gaussian
  functions model. We assume the effective survey area and selection
  function reported for the DR3 quasar sample \citep{dr3lumfunc}.

  \subsection{Construction of the Simulated Sample}

  \label{s-simconst}

  We first drew a random value of $N_{\Omega}$ quasars from a binomial
  distribution with probability of success $\Omega / 4 \pi = 0.0393$
  and number of trials $N = 3 \times 10^5$. Here, $\Omega = 1622\ {\rm
  deg}^2$ is the effective sky area for our simulated survey, and we
  chose the total number of quasars to be $N = 3 \times 10^5$ in order
  to ultimately produce a value of $n \sim 1300$ observed sources,
  after accounting for the SDSS selection function. This first step of
  drawing from a binomial distribution simulates a subset of
  $N_{\Omega} \sim 1.2 \times 10^4$ sources from $N$ total sources
  randomly falling within an area $\Omega$ on the sky. For simplicity,
  in this simulation we ignore the effect of obscuration on the
  observed quasar population. While our choice of $N = 3 \times 10^5$
  produces a much smaller sample than the actual sample of $n \sim 1.5
  \times 10^4$ quasars from the SDSS DR3 luminosity function work
  \citep{dr3lumfunc}, we chose to work with this smaller sample to
  illustrate the effectiveness of our method on more moderate sample
  sizes.

  For each of these $N_{\Omega} \sim 1.2 \times 10^4$ sources, we
  simulated values of $L$ and $z$. We first simulated values of $\log
  z$ from a marginal distribution of the form
  \begin{equation}
    f(\log z) = \frac{4\Gamma(a + b)}{\Gamma(a) \Gamma(b)} \frac{\exp(a \zeta^*)}
               {\left(1 + \exp(\zeta^*) \right)^{a+b}}, \label{eq-zmarg}
  \end{equation}
  where $\zeta^* = 4 (\log z - 0.4)$. The parameters $a = 1.25$ and $b =
  2.5$ were chosen to give an observed redshift distribution similar to
  that seen for SDSS DR3 quasars \citep[e.g.,][]{dr3lumfunc}. Values
  of $\log z$ are easily drawn from Equation (\ref{eq-zmarg}) by first
  drawing $x^* \sim {\rm Beta}(a,b)$, and then setting $\log z = {\rm
  logit}(x^*) / 4 + 0.4$; here, ${\rm Beta}(a,b)$ is a beta
  probability density, and ${\rm logit}(x) = \ln (x / (1 - x))$ is the
  logit function.

  For each simulated value of $z$, we simulated a value of $L$ using a
  similar functional form. The conditional distribution
  of $\log L$ given $z$ is
  \begin{eqnarray}
    f(\log L|z) & = & \frac{\Gamma(\alpha(z) + \beta(z))}{\Gamma(\alpha(z)) 
      \Gamma(\beta(z))} \frac{(L / L^*(z))^{\alpha(z) / \ln 10}}
    {\left[1 + (L / L^*(z))^{1 / \ln 10} \right]^{\alpha(z)+\beta(z)}}
    \label{eq-mcond} \\
    \alpha(z) & = & 6 + \log z \\
    \beta(z) & = & 9 + 2 \log z \\
    L^*(z) & = & 10^{45} z^2,
  \end{eqnarray}
  where $L^*(z)$ approximately marks the location of the peak in
  $f(\log L|z)$, $t(z)$ is the age of the universe in Gyr at redshift
  $z$, $\alpha(z)$ is the slope of $\log f(\log L|z)$ for $L \lesssim
  L^*(z)$, and $\beta(z)$ is the slope of $\log f(\log L|z)$ for $L
  \gtrsim L^*(z)$. In this simulated `universe', both the peak and
  logarithmic slopes of the LF evolve. The form of the luminosity
  function assumed by Equation (\ref{eq-mcond}) is similar to the
  double power-law form commonly used in the quasar LF literature, but
  has a more gradual transition between the two limiting slopes.

\begin{figure}
  \begin{center}
    \scalebox{0.7}{\rotatebox{90}{\plotone{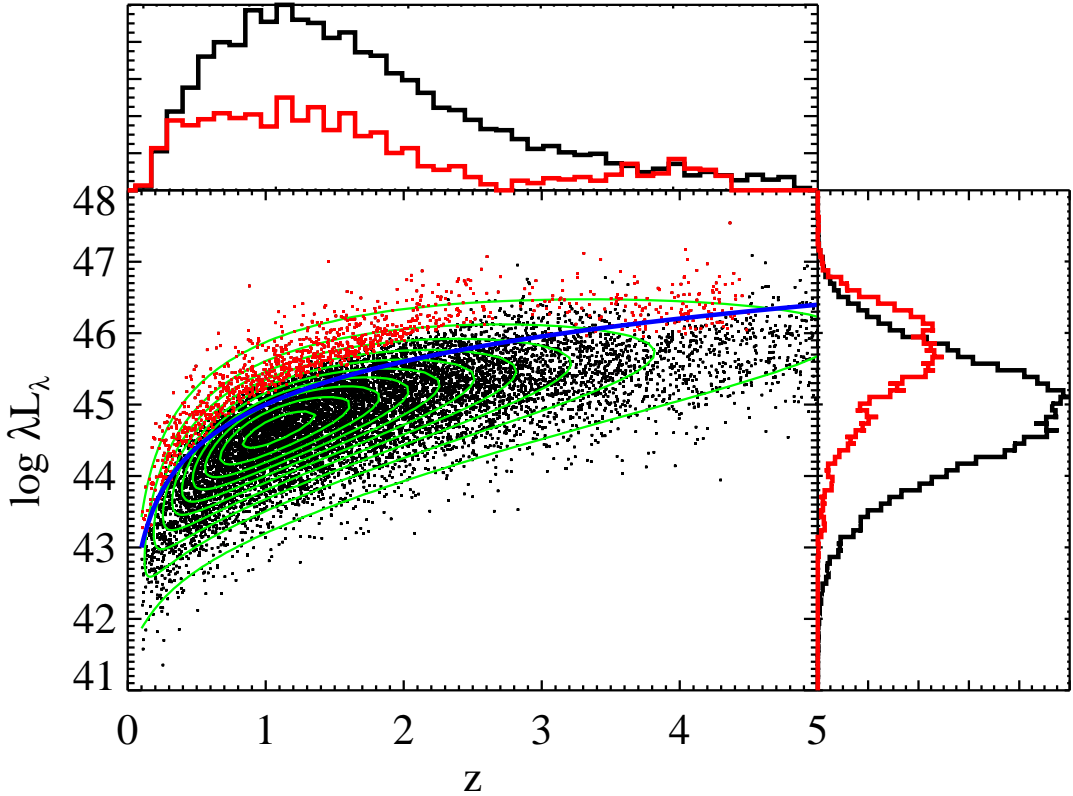}}}
    \caption{The distribution of $L$ and $z$ for the simulated sample
    described in \S~\ref{s-simconst}. Red dots denote sources included
    in the sample, black dots denote sources not included in the
    sample, the blue line denotes $L^*$ as a function of $z$, and the
    green contours display the 2-d luminosity function. Also shown are
    histograms of the marginal distributions of $\log L$ and $z$, for
    all simulated objects (black histogram) and only the detected ones
    (red histogram). For clarity, the histogram of the detected
    sources has been forced to peak at a value equal to half of the
    peak of the histogram of all objects.
    \label{f-simdist}}
  \end{center}
\end{figure}

  After using Equations (\ref{eq-zmarg}) and (\ref{eq-mcond}) to
  generate random values of $L$ and $z$, we simulated the effects at a
  selection function. We randomly kept each source for $z < 4.5$,
  where the probability of including a source given its luminosity and
  redshift was taken to be the SDSS DR3 Quasar selection function, as
  reported by \citet{dr3lumfunc}. After running our simulated sample
  through the selection function, we were left with a sample of $n
  \sim 1300$ sources. Therefore, our simulated survey is only able to
  detect $\sim 0.4\%$ of the $N = 3 \times 10^5$ total quasars in our
  simulated `universe'. The distributions of $L$ and $z$ are shown in
  Figure \ref{f-simdist} for both the detected sources and the full
  sample. As can be seen, the majority of sources are missed by our
  simulated survey.

\begin{figure}
  \begin{center}
    \includegraphics[scale=0.7,angle=90]{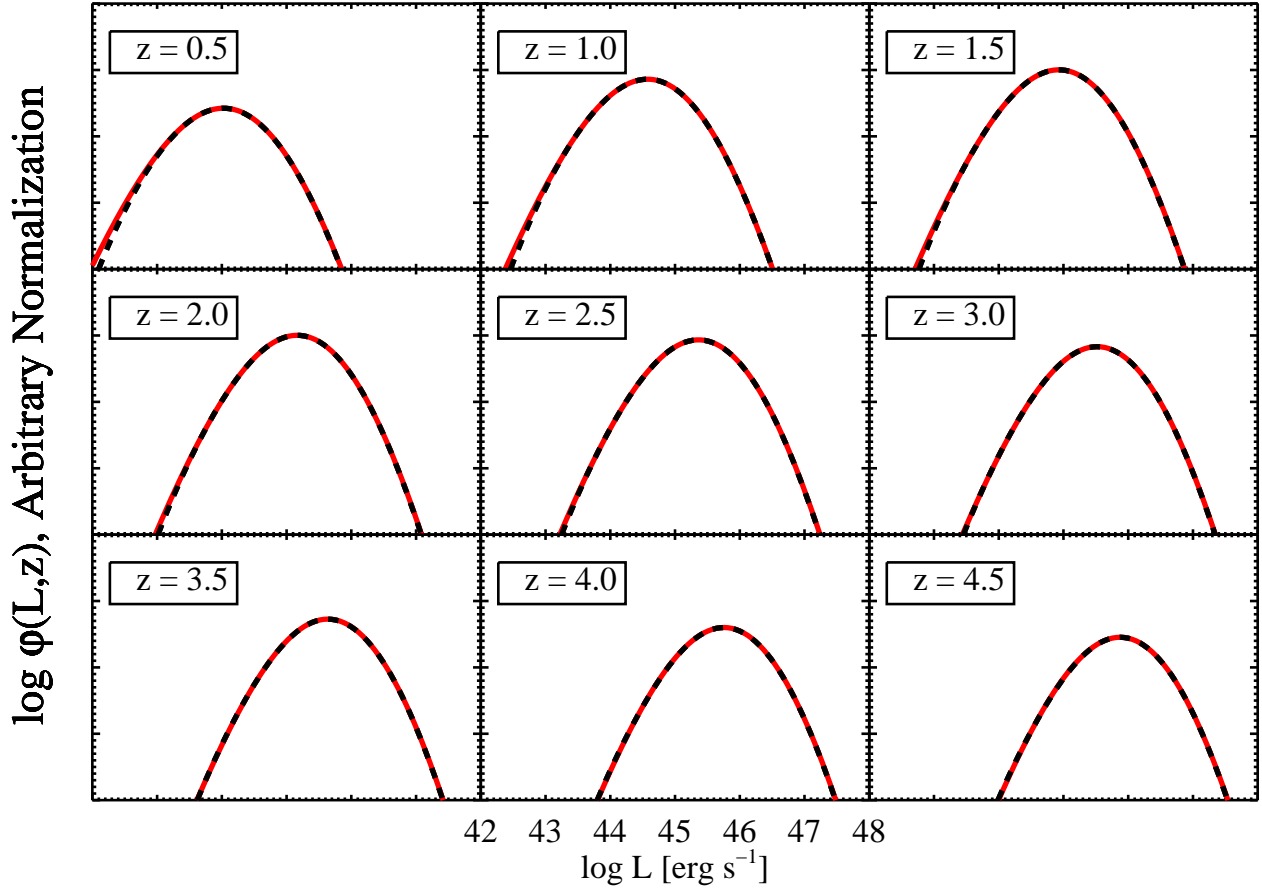}
    \caption{The true LF (solid red line) at several values of $z$,
    and the best $K=4$ Gaussian function fit (dashed black line). In
    this case, approximating the LF with four 2-dimensional Gaussian
    functions provides an excellent fit.\label{f-truebhmf}}.
  \end{center}
\end{figure}

  The joint probability distribution of $L$ and $z$ is $f(L,z) =
  f(L|z) f(z)$, and therefore Equations (\ref{eq-zmarg}) and
  (\ref{eq-mcond}) imply that the true LF for our simulated sample is
  \begin{equation}
    \phi_0(L,z) = \frac{N}{z L (\ln 10)^2} \left(\frac{dV}{dz}\right)^{-1} 
    f(\log L|z) f(\log z)
    \label{eq-bhmftrue}
  \end{equation}
  Figure \ref{f-truebhmf} shows $\phi_0(L,z)$ at several
  redshifts. Also shown in Figure \ref{f-truebhmf} is the best fit for
  a mixture of $K = 4$ Gaussian functions. Despite the fact that
  $\phi_0(L,z)$ has a rather complicated parametric form, a mixture
  of four Gaussian functions is sufficient to achieve an excellent
  approximation to $\phi_0(L,z)$; in fact, the mixture of four
  Gaussian functions approximation is indistinguishable from the true
  LF.

  \subsection{Performing Statistical Inference on the LF with the MCMC Output}

  \label{s-simmcmc}

  We performed the MHA algorithm described in \S~\ref{s-mha} to obtain
  random draws from the posterior probability distribution for our
  this simulated sample, assuming the Gaussian mixture model described
  in \S~\ref{s-smodel}. We performed $10^4$ iterations of burn-in, and
  then ran the Markov chains for $3 \times 10^4$ more iterations. We
  ran 20 chains simultaneously in order to monitor convergence
  \citep[e.g., see][]{gelman04} and explore possible multimodality in
  the posterior. We saved the values of $\theta$ for the Markov chains
  after the initial $10^4$ burn-in iterations, and, after removing
  divergent chains with $N < 0$ we were left with $\sim 8 \times 10^4$
  random draws from the posterior distribution, $p(\theta, N|L_{obs},
  z_{obs})$.

  The output from the MCMC can be used to perform statistical
  inference on the LF. Denote the $T \sim 8 \times 10^4$ random draws
  of $\theta$ and $N$ obtained via the MHA as $\theta^1, \ldots,
  \theta^T$ and $N^1, \ldots, N^T$, respectively. The individual
  values of $(\theta^t,N^t)$ can then be used to construct histograms
  as estimates of the posterior distribution for each parameter. For
  each random draw of $\theta$ and $N$, we can also calculate a random
  draw of $\phi(L,z)$ from its posterior distribution. In particular,
  the $t^{\rm th}$ random draw of the LF under the mixture of normals
  model, denoted as $\phi^t(L,z)$, is calculated by inserting
  $\theta^t$ and $N^t$ into Equation (\ref{eq-mixlf}). The $T$ values
  of $\phi^t(L,z)$ can then be used to estimate the posterior
  distribution of $\phi(L,z)$ for any given value of $L$ and
  $z$. Furthermore, random draws from the posterior for quantities
  that are computed directly from the LF, such as the location of its
  peak as a function of $z$, are obtained simply by computing the
  quantity of interest from each of the $T$ values of $\phi^t(L,z)$.

\begin{figure}
  \begin{center}
    \scalebox{0.7}{\rotatebox{90}{\plotone{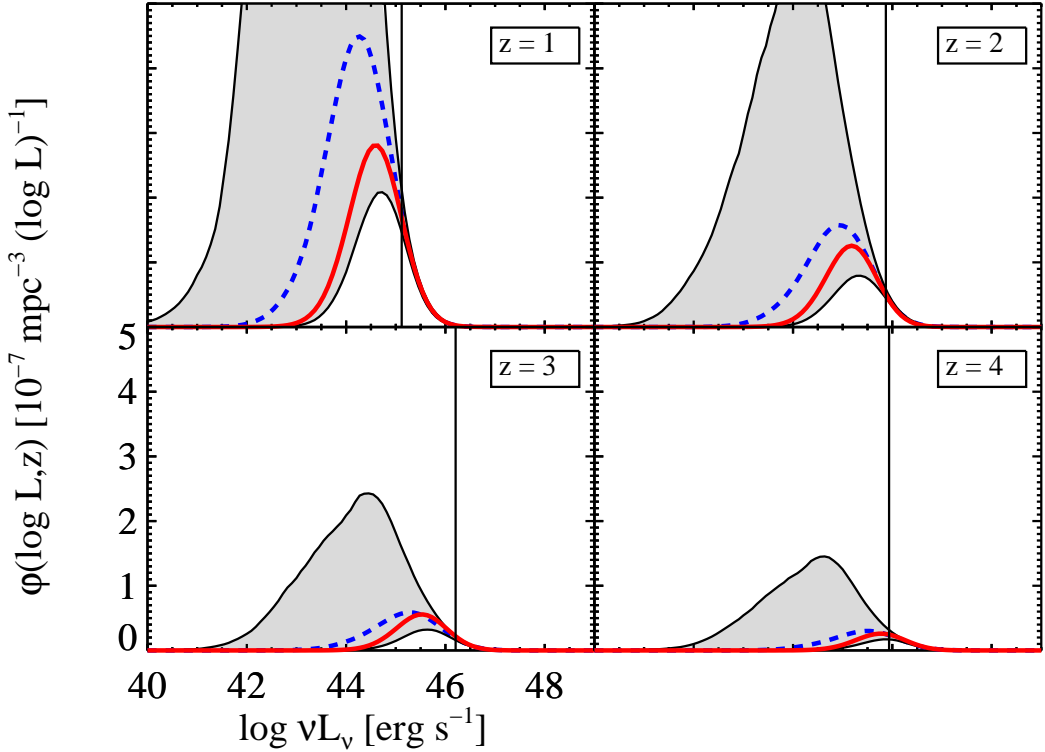}}}
    \caption{The true LF (solid red line) at several redshifts for the
      simulated sample described in \S~\ref{s-simconst}. The axis
      labels are the same for all panels, but for clarity we only
      label the bottom left panel. Also shown is the posterior median
      estimate of the LF based on the mixture of Gaussian functions
      model (dashed blue line), the region containing $90\%$ of the
      posterior probability (shaded region). The bayesian mixture of
      Gaussian functions model is able to accurately constrain the LF,
      even below the survey detection limit. \label{f-philin}}
  \end{center}
\end{figure}

\begin{figure}
  \begin{center}
    \scalebox{0.7}{\rotatebox{90}{\plotone{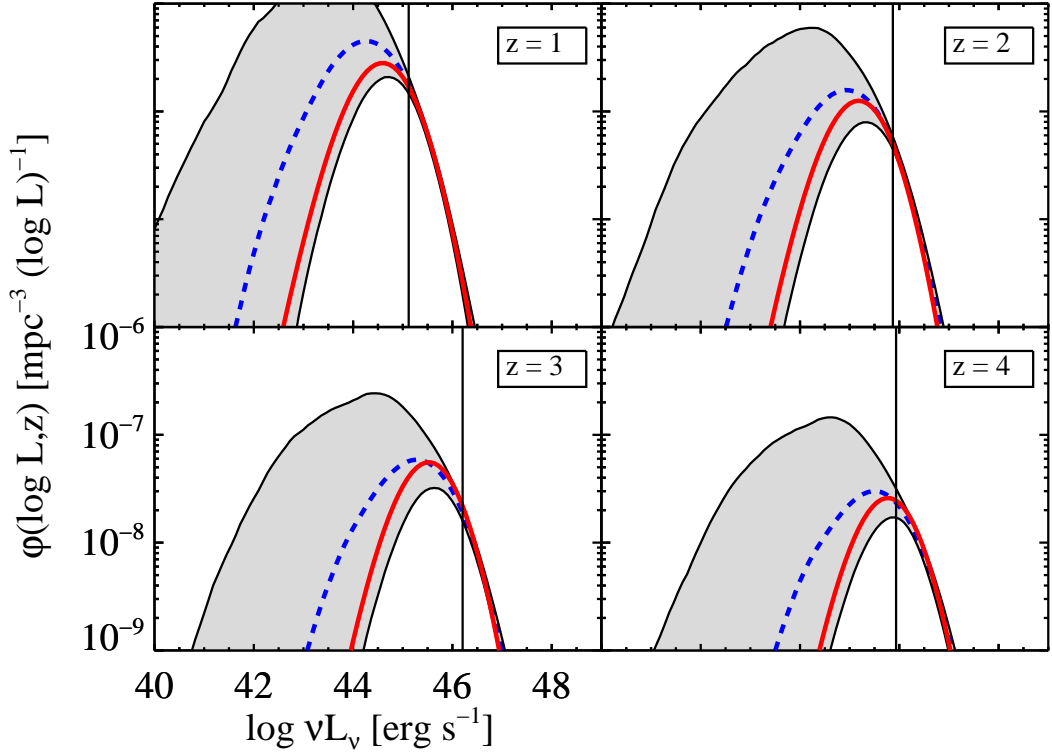}}}
    \caption{Same as Figure \ref{f-philin}, but shown with a
      logarithmic stretch.\label{f-philog}}
  \end{center} 
\end{figure}

  In Figures \ref{f-philin} and \ref{f-philog} we show $\phi(\log
  L,z)$ at several different redshifts, on both a linear scale and a
  logarithmic scale. In general, we find it easier to work with
  $\phi(\log L,z) = \ln 10 L \phi(L,z)$, as $\phi(\log L,z)$ can span
  several orders of magnitude in $L$. Figures \ref{f-philin} and
  \ref{f-philog} show the true value of the LF, $\phi_0(\log L,z)$,
  the best-fit estimate of $\phi(\log L,z)$ based on the mixture of
  Gaussian functions model, and the regions containing $90\%$ of the
  posterior probability. Here, as well as throughout this work, we
  will consider the posterior median of any quantity to be the
  `best-fit' for that quantity. In addition, in this work we will
  report errors at the $90\%$ level, and therefore the regions
  containing $90\%$ of the posterior probability can be loosely
  interpreted as asymmetric error bars of length $\approx
  1.65\sigma$. The region containing $90\%$ of the probability for
  $\phi(\log L,z)$ is easily estimated from the MCMC output by finding
  the values of $t_1$ and $t_2$ such that $90\%$ of the values of
  $\phi^1(\log L,z), \ldots, \phi^T(\log L,z)$ have $\phi^{t_1}(\log
  L,z) < \phi^t (\log L,z) < \phi^{t_2}(\log L,z)$. As can be seen,
  the true value of $\phi(\log L,z)$ is contained within the $90\%$
  probability region for all almost values of $L$, even those below
  the survey detection limit.

\begin{figure}
  \begin{center}
    \includegraphics[scale=0.33,angle=90]{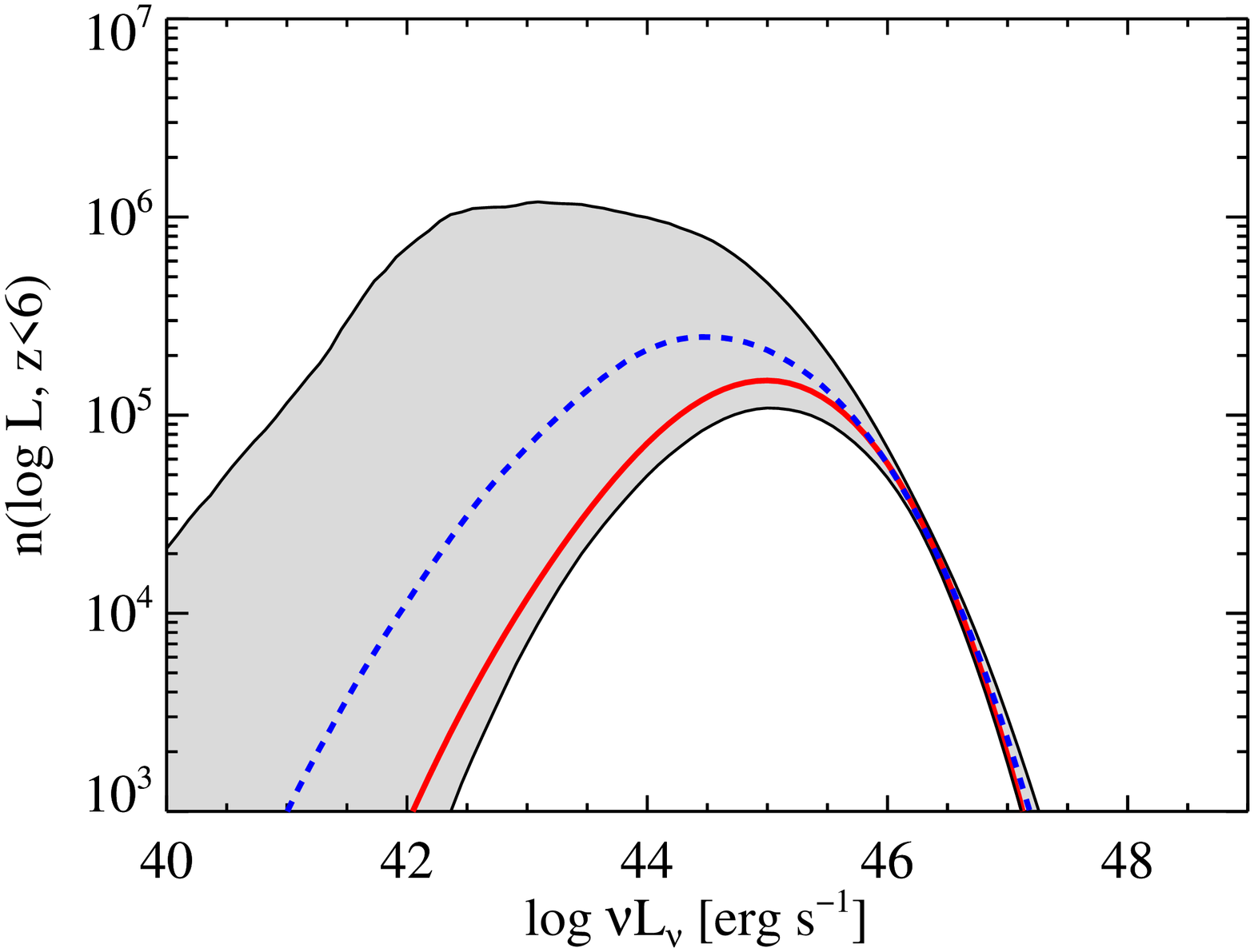}
    \includegraphics[scale=0.33,angle=90]{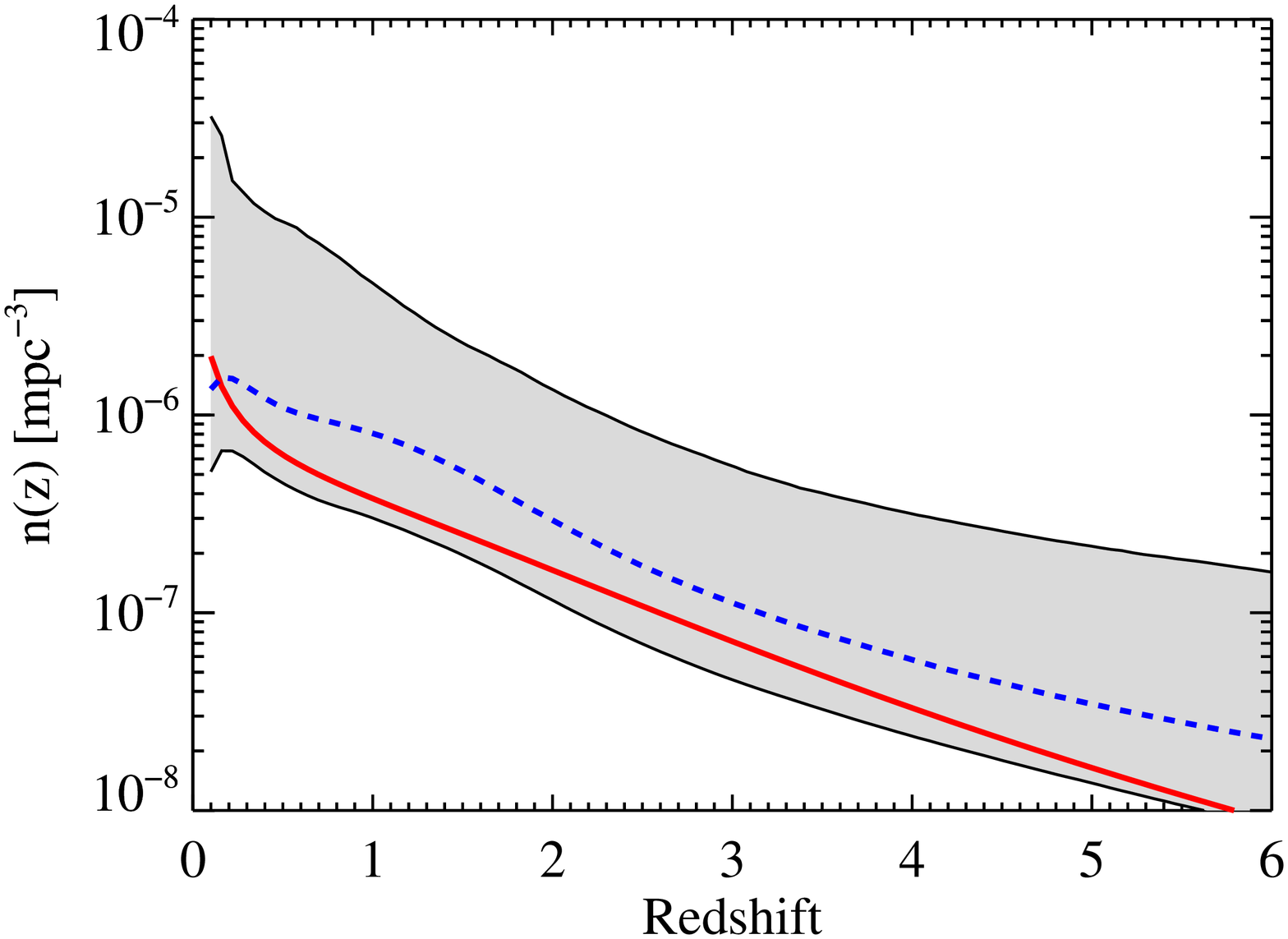}
    \caption{The integrated $z < 6$ quasar number density (number per
    $\log L$ interval, left) and the quasar comoving quasar number
    density as a function of $z$ (number per ${\rm mpc}^3$, right) for
    the simulated sample described in \S~\ref{s-simconst}. As with
    Figure \ref{f-philin}, the solid red line denotes the true value
    for the simulation, the dashed blue line denotes the posterior
    median for the mixture of Gaussian functions model, and the shaded
    region contain $90\%$ of the posterior probability. The posterior
    median provides a good fit to the true values, and the
    uncertainties derived from the MCMC algorithm based on the
    Gaussian mixture model are able to accurately constrain the true
    values of these quantities, despite the flux
    limit.\label{f-mzmarg}}
  \end{center}
\end{figure}

  Figure \ref{f-mzmarg} compares the true integrated $z < 6$ number
  distribution of $\log L$, $n(\log L, z < 6)$, with the mixture of
  Gaussian functions estimate.  The quantity $n(\log L, z<6) d\log L$
  gives the number of quasars at $z < 6$ with black hole masses
  between $\log L$ and $\log L + d\log L$. It is calculated as
  \begin{equation}
    n(\log L, z < 6) = \int_0^6 \phi(\log L,z) \left(\frac{dV}{dz}\right)\ dz,
    \label{eq-mnumbdist}
  \end{equation}
  which, for the mixture of normals model, is
  \begin{eqnarray}
    n(\log L, z < z_0) & = & N \sum_{k=1}^K \pi_k N(\log L|\mu_{l,k},\sigma^2_{l,k}) 
    \Phi\left[\frac{\log z_0 - E(\log z|L,k)}{\sqrt{Var(\log z|L,k)}} \right] 
    \label{eq-mnumbdist_mix} \\
    E(\log z|L,k) & = & \mu_{z,k} + \frac{\sigma_{lz,k}}{\sigma^2_{z,k}} 
    \left(\log L - \mu_{l,k} \right) \\
    Var(\log z|L,k) & = & \sigma^2_{z,k} - \frac{\sigma^2_{lz,k}}{\sigma^2_{z,k}}.
  \end{eqnarray}
  Here, $\Phi(\cdot)$ is the cumulative distribution function for the
  standard normal density. Similar to Figures \ref{f-philin} and
  \ref{f-philog}, the true value of $n(\log L,z < 6)$ is contained
  within the $90\%$ probability region for all values of $L$, even
  those below the survey detection limit.

  In addition, in Figure \ref{f-mzmarg} we show the comoving number
  density of broad line AGN as a function of redshift, $n(z)$. This is
  obtained by integrating $\phi(L,z)$ over all possible values of
  $L$. For the mixture of normals model, this becomes
  \begin{equation}
    n(z) = N \left(\frac{dV}{dz}\right)^{-1} \sum_{k=1}^K \pi_k p(z|k),
    \label{eq-comov_nz}
  \end{equation}
  where the marginal distribution of $z|k$ is
  \begin{equation}
    p(z|k) = \frac{1}{z \ln 10 \sqrt{2\pi \sigma^2_{z,k}}} \exp \left\{-\frac{1}{2}
    \left(\frac{\log z - \mu_{z,k}}{\sigma_{z,k}} \right)^2 \right\}.
    \label{eq-zmargmix}
  \end{equation}
  As before, the true value of $n(z)$ is contained within the $90\%$
  probability region, despite the fact that the integration extends
  over \emph{all} $L$, even those below the detection limit. The wider
  confidence regions reflect additional uncertainty in $n(z)$
  resulting from integration over those $L$ below the detection
  limit. In particular, the term $dV / dz$ becomes small at low
  redshift, making the estimate of $n(z)$ more unstable as $z
  \rightarrow 0$, and thus inflating the uncertainties at low $z$.

\begin{figure}
  \begin{center}
    \scalebox{0.7}{\rotatebox{90}{\plotone{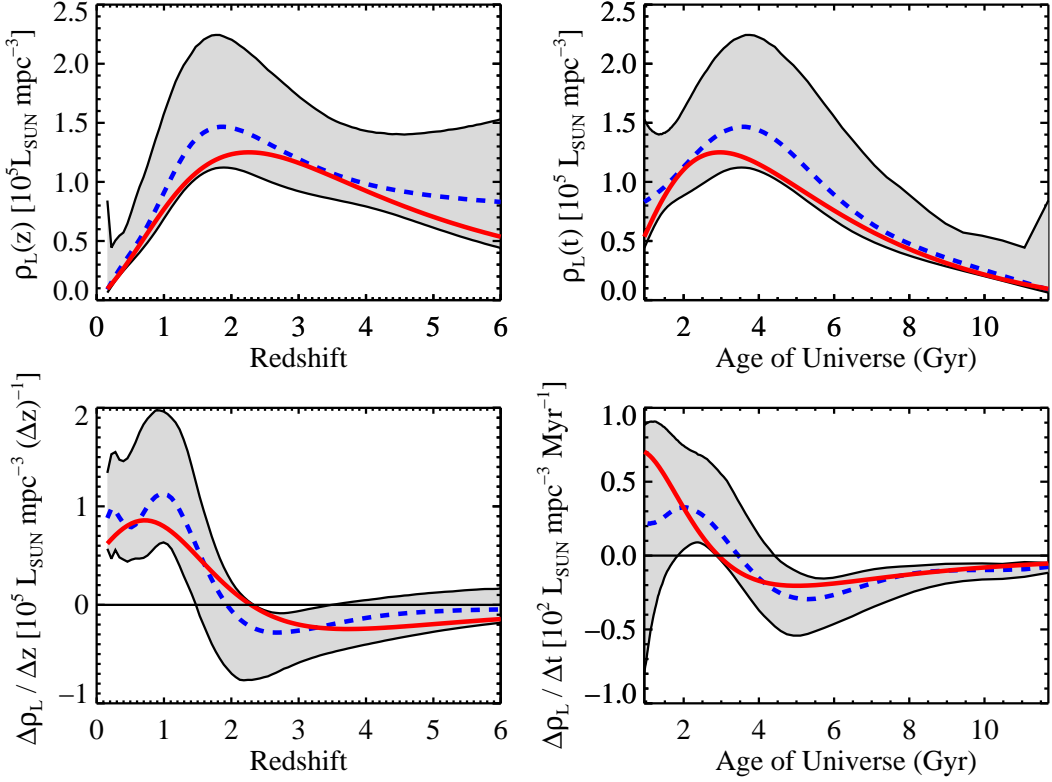}}}
    \caption{Comoving quasar luminosity density (top two panels) and
    its derivative (bottom two panels), shown as a function of
    redshift (left two panels) and cosmic age (right two panels) for
    the simulated sample described in \S~\ref{s-simconst}. The
    plotting symbols are the same as in Figure \ref{f-mzmarg}. As in
    the previous figures, the Gaussian mixture model is able to
    provide an accurate fit to the true values of $\rho_{L}(z)$, and
    the bayesian MCMC approach is able to provide accurate constraints
    on $\rho_{L}(z)$ and $d\rho_{L} / dz$, despite the fact that the
    integral used for calculating these quanties extends below the
    survey detection limit. \label{f-rhoz}}
  \end{center} 
\end{figure}

  Two other potentially useful quantities are the comoving luminosity
  density for quasars, $\rho_L(z)$, and its derivative. The comoving
  quasar luminosity density is given by $\rho_L(z) = \int_0^{\infty} L
  \phi(L,z)\ dL$. For the mixture of Gaussian functions model it may
  be shown that
  \begin{eqnarray}
    \rho_L(z) & = & N \left(\frac{dV}{dz}\right)^{-1} \sum_{k=1}^{K} \pi_k p(z|k) 
    \exp \left \{ \ln 10 E(\log L|z,k) + \frac{(\ln 10)^2}{2} Var(\log L|z,k) \right \}
    \label{eq-rhoz} \\
    E(\log L|z,k) & = & \mu_{l,k} + \frac{\sigma_{lz,k}}{\sigma^2_{z,k}} 
    \left( \log z - \mu_{z,k} \right) \\
    Var(\log L|z,k) & = & \sigma^2_{l,k} - \frac{\sigma^2_{lz,k}}{\sigma^2_{z,k}},
  \end{eqnarray}
  where $p(z|k)$ is given by Equation (\ref{eq-zmargmix}). We
  calculate the derivative of $\rho_L(z)$ numerically. Figure
  \ref{f-rhoz} compares the true values of $\rho_L(z)$ and its
  derivative with the posterior distribution for $\rho_L(z)$ inferred
  from the mixture model, both as a function of $z$ and the age of the
  universe at redshift $z$, $t(z)$. Comparison with Figure
  \ref{f-mzmarg} reveals that the comoving quasar luminosity density,
  $\rho_L(z)$, is a better constrained quantity than the comoving
  quasar number density, $n(z)$.  Furthermore, $n(z)$ appears to peak
  much later than $\rho_L(z)$. In addition, we can correctly infer
  that the comoving quasar luminosity density reaches it point of
  fastest growth at $t(z) \sim 2$ Gyr, and its point of fastest
  decline at $t(z) \sim 5$ Gyr.

\begin{figure}
  \begin{center}
    \includegraphics[scale=0.33,angle=90]{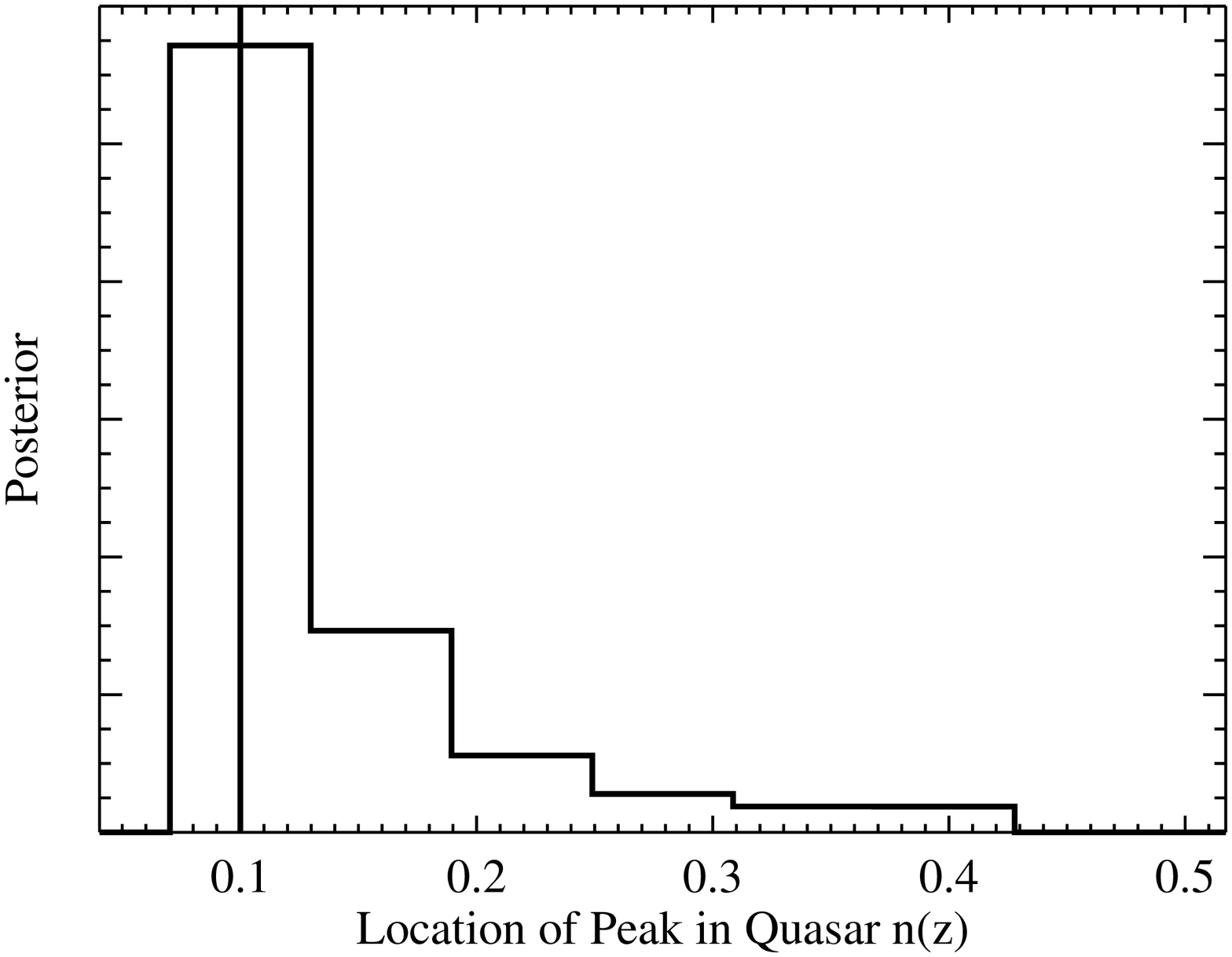}
    \includegraphics[scale=0.33,angle=90]{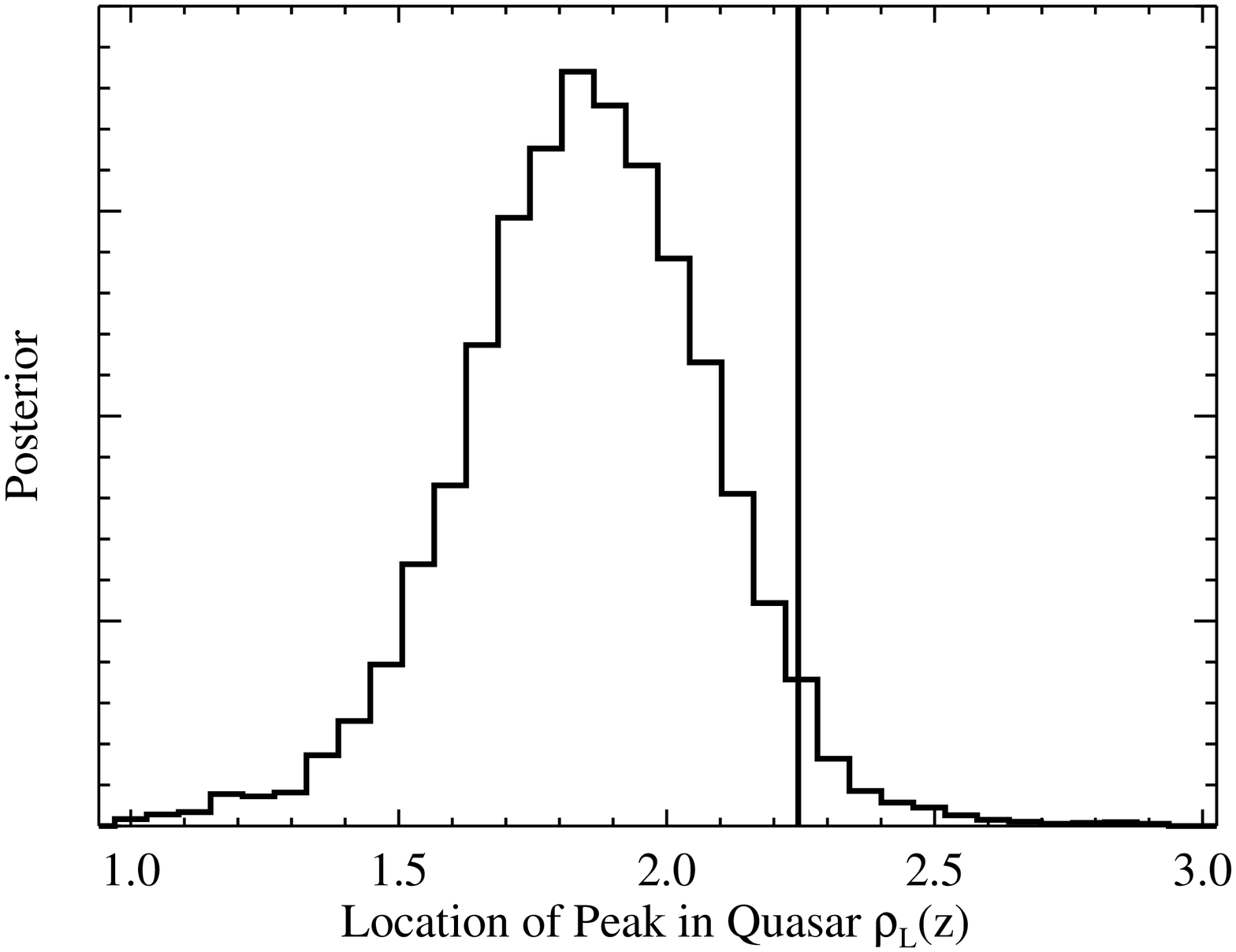}
    \caption{Posterior distribution for the redshift location of the
    peak in the comoving number density of quasars (left) and the peak
    in the comoving quasar luminosity density (right) for the
    simulated sample described in \S~\ref{s-simconst}. For clarity we
    only show the posterior distribution for the peak in $n(z)$ at $z
    > 0.5$, since values of the peak at $z < 0.5$ arise because the
    term $(dV / dz)^{-1}$ becomes very large at low $z$. The vertical
    lines denote the true values. The posterior distribution inferred
    from the MCMC output is able to accurately constrain the true
    values of the argumentative maximum in $n(z)$ and
    $\rho_{L}(z)$.\label{f-peaks}}
  \end{center}
\end{figure}

\begin{figure}
  \begin{center}
    \includegraphics[scale=0.33,angle=90]{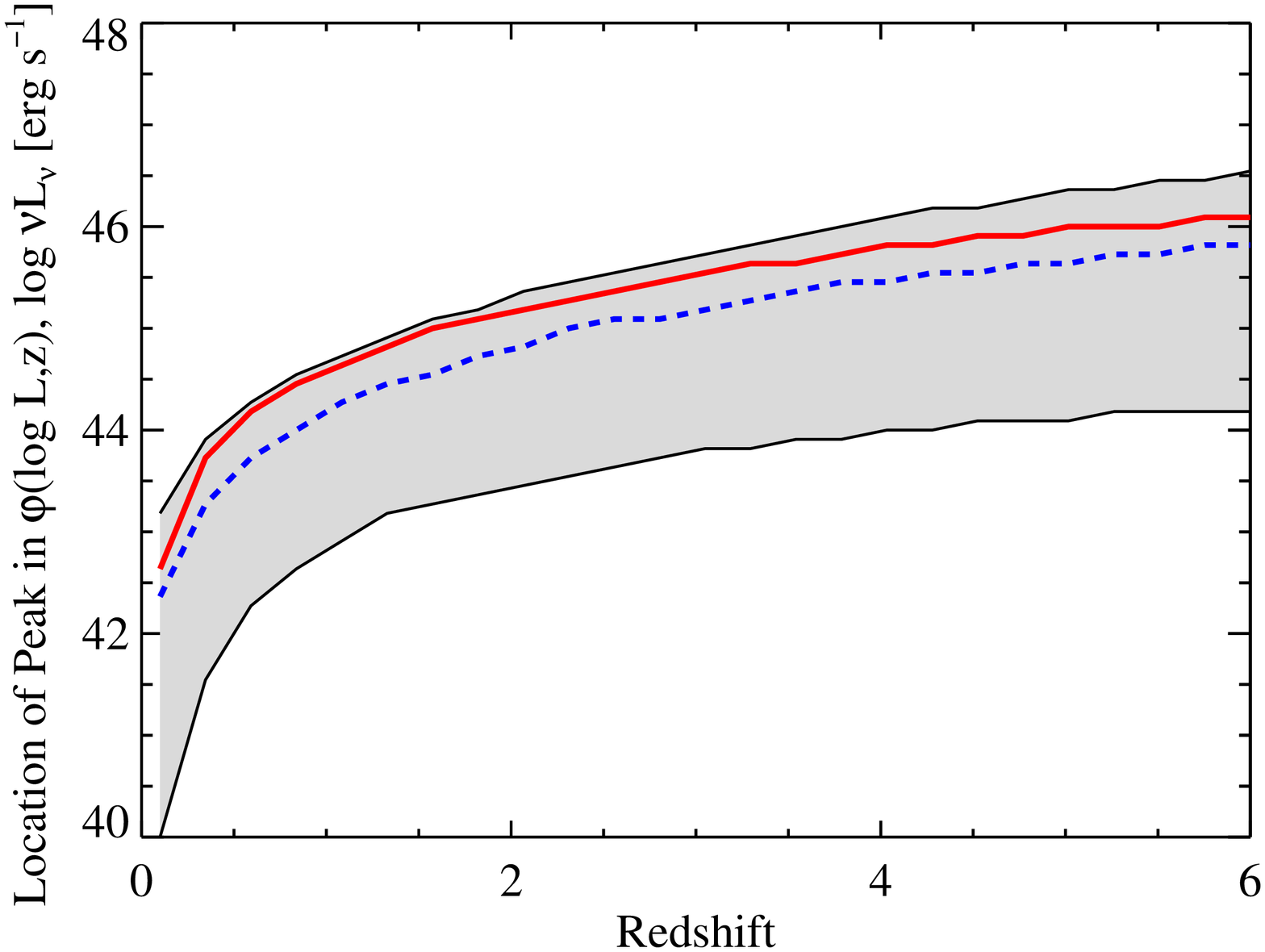}
    \includegraphics[scale=0.33,angle=90]{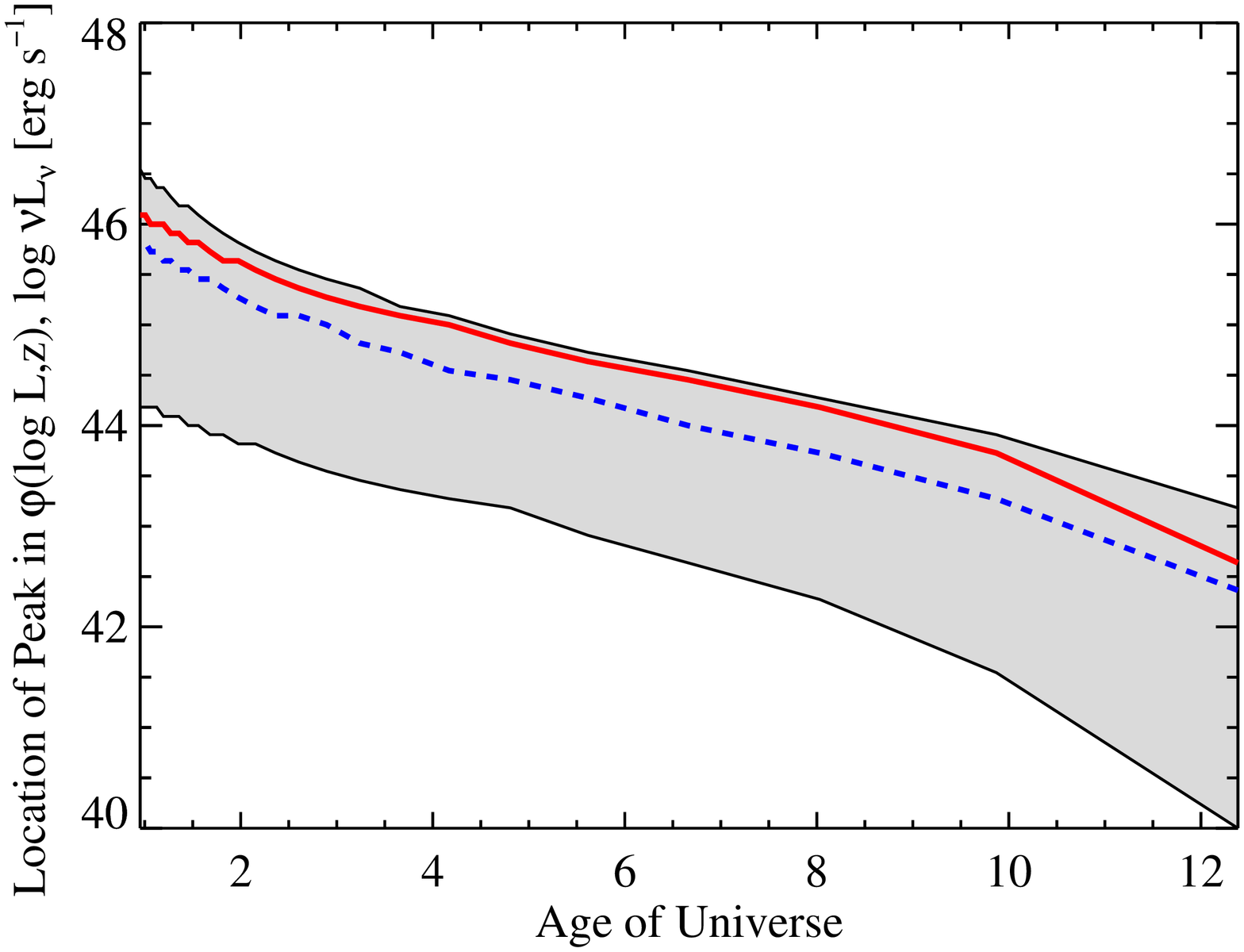}
    \caption{Location of the peak in the LF as a function of $z$
    (left) and cosmic age (right) for the simulated sample described
    in \S~\ref{s-simconst}. The plot symbols are the same is in Figure
    \ref{f-mzmarg}. In general the posterior median of the Gaussian
    mixture model provides a good estimate of the true peak locations,
    although the uncertainty is high due to the survey flux
    limit. However, it is clear from these plots that the location of
    the peak in $\phi(L,z)$ evolves.\label{f-mpeakevol}}
  \end{center}
\end{figure}

  Figure \ref{f-peaks} quantifies the suggestion that $n(z)$ peaks
  later than $\rho_L(z)$ by displaying the posterior distribution for
  the location of the respective peaks in $n(z)$ and $\rho_L(z)$. We
  can still constrain the peak in $n(z)$ to be at $z \lesssim 0.5$. In
  contrast, the location of the peak in $\rho_L(z)$ is constrained to
  occur earlier at $1 \lesssim z \lesssim 3$. This is a consequence of
  the fact that while there were more quasars per comoving volume
  element in our simulated universe at $z \lesssim 0.5$, their
  luminosities were much higher at higher redshift. This evolution in
  characteristic $L$ is quantified in Figure \ref{f-mpeakevol}, which
  summarizes the posterior distribution for the location of the peak
  in $\phi(\log L,z)$ as a function of redshift and $t(z)$. As can be
  seen, the location of the peak in the LF shows a clear trend of
  increasing `characteristic' $L$ with increasing $z$, although there
  is considerable uncertainty on the actual value of the location of
  the peak.

  \subsection{Using the MCMC Output to Evaluate the LF Fit}

  \label{s-postcheck}

  Throughout this section we have been analyzing the MCMC results by
  comparing to the true LF. However, in practice we do not have
  access to the true LF, and thus a method is needed for assessing
  the quality of the fit. The statistical model may be checked using a
  technique known as posterior predictive checking
  \citep[e.g.,][]{rubin81,rubin84,gelman98}. Here, the basic idea is
  to use each of the MCMC outputs to simulate a new random observed
  data set. The distributions of the simulated observed data sets are
  then compared to the true observed data in order to assess whether
  the statistical model gives an accurate representation of the
  observed data. It is important to construct simulated data sets
  for each of the MCMC draws in order to incorporate our uncertainty
  in the model parameters.

  For each value of $N^t$ and $\theta^t$ obtained from the MCMC
  output, a simulated data set of $(l^t_{obs}, z^t_{obs})$ may be
  obtained through a similar procedure to that described in
  \S~\ref{s-simconst}. First, one draws a value of $N_{\Omega}^t$ from
  a binomial distribution with $N^t$ trials and probability of
  `success' $p = \Omega / 4\pi$. Then, one draws $N_{\Omega}^t$ values
  of $L^t$ and $z^t$ from $p(L,z|\theta^t)$.

  For our model, $p(\log L,\log z|\theta^t)$ is a mixture of normal
  densities, and one needs to employ a two-step process in order to
  simulate a random value from $p(\log L,\log z|\theta^t)$. First, one
  needs to randomly assign the $i^{\rm th}$ data point to one of the
  Gaussian distributions. Since $\pi_k$ gives the probability that a
  data point will be drawn from the $k^{\rm th}$ Gaussian
  distribution, one first needs to simulate a random vector ${\bf
  G}^t_i$ from a multinomial distribution with one trial and
  probability of success for the $k^{\rm th}$ class $\pi^t_k$; i.e.,
  first draw ${\bf G}^t_i \sim {\rm
  Multinom}(1,\pi^t_1,\ldots,\pi^t_K)$. The vector ${\bf G}^t_i$ gives
  the class membership for the $i^{\rm th}$ data point, where
  $G^t_{ik} = 1$ if the $i^{\rm th}$ data point comes from the $k^{\rm
  th}$ Gaussian, and $G^t_{ij} = 0$ if $j \neq k$. Then, given
  $G^t_{ik} = 1$, one then simulates a value of $(\log L_i^t,\log
  z_i^t)$ from a 2-dimensional Gaussian distribution with mean
  $\mu_k^t$ and covariance matrix $\Sigma^t_k$. This is repeated for
  all $N^t_{\Omega}$ sources, leaving one with a random sample $(\log
  L^t,\log z^t) \sim p(\log L,\log z|\theta^t)$.

  A random draw from ${\rm Multinom}(1,\pi_1,\ldots,\pi_K)$, may be
  obtained as a sequence of binomial random draws. First, draw $n'_1
  \sim {\rm Binomial}(1,\pi_1)$. If $n'_1 = 1$, then assign the data
  point to the first Gaussian distribution, i.e., set $G_{i1} = 1$. If
  $n'_1=0$, then draw $n'_2 \sim {\rm Binomial}(1,\pi_2 / \sum_{k=2}^K
  \pi_k)$. If $n'_2 = 1$, then assign the data point to the second
  Gaussian distribution, i.e., set $G_{i2} = 1$. If $n'_2 = 0$, then
  the process is repeated for the remaining Gaussian distribution as
  follows. For $j = 3, \ldots, K-1$, sequentially draw $n'_j \sim {\rm
  Binomial}(1,\pi_j / \sum_{k=j}^{K} \pi_k)$. If at any time $n'_j =
  1$, then stop the process and assign the data point to the $j^{\rm
  th}$ Gaussian distribution. Otherwise, if none of the $n'_j = 1$,
  then assign the data point to the $K^{\rm th}$ Gaussian
  distribution.

  Once one obtains a random draw of $(L^t,z^t)$, randomly `observe'
  these sources, where the probability of including a source given
  $L^t_i$ and $z^t_i$ is given by the selection function. This will
  leave one with a simulated observed data set, $(L_{obs}^t,
  z^t_{obs})$. This process is repeated for all $T$ values of $N^t$
  and $\theta^t$ obtained from the MCMC output, leaving one with $T$
  simulated data sets of $(L_{obs}^t, z^t_{obs})$. One can then
  compare the distribution of the simulated data sets with the true
  values of $L_{obs},$ and $z_{obs}$ to test the statistical model for
  any inconsistencies.

\begin{figure}
  \begin{center}
    \scalebox{0.7}{\rotatebox{90}{\plotone{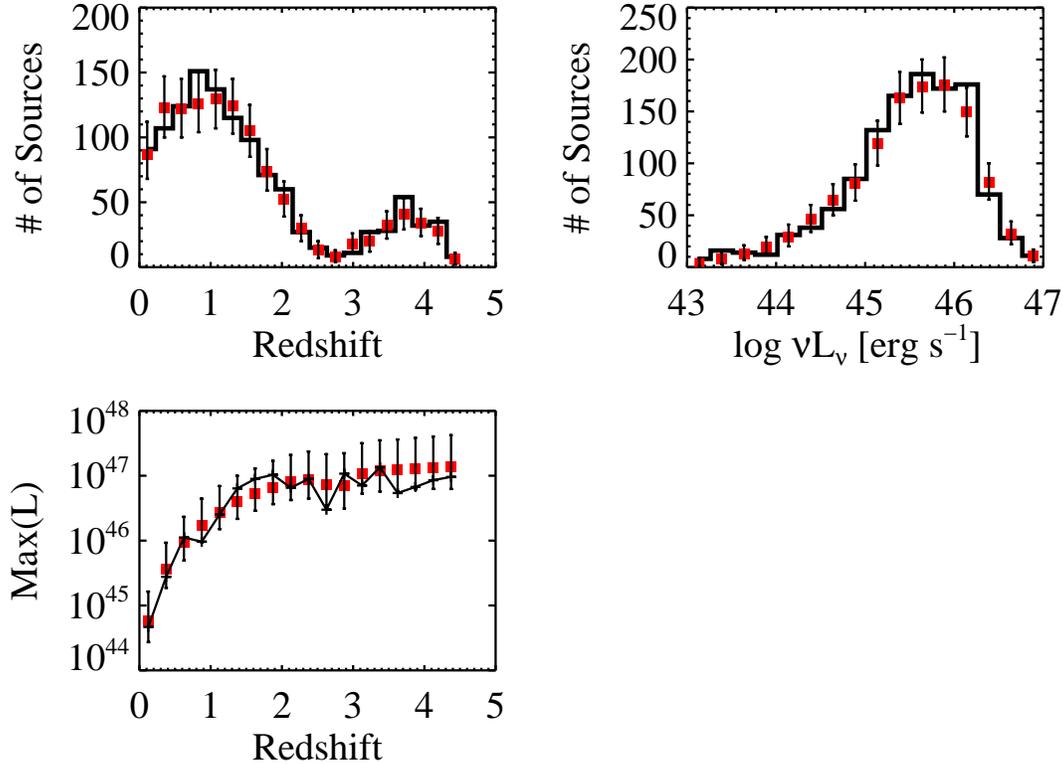}}}
    \caption{Posterior predictive check for the Gaussian mixture model
    (see \S~\ref{s-postcheck}). The histograms show the actual
    distributions of $L_{obs}$ and $z_{obs}$, the red squares denote
    the posterior medians for the number of sources in each respective
    bin, and the error bars contain the inner $90\%$ of the histogram
    values for the samples simulated from the posterior. Also shown is
    a plot of the maximum observed luminosity as a function of $z$ for
    the simulated samples, the red squares mark the value from the
    actual sample used in the fit, and the error bars contain $90\%$
    of the values simulated from the posterior distribution. The
    mixture of Gaussian functions model is able to provide an accurate
    prediction of the observed distribution of luminosity, redshift,
    and line widths, and thus there is not any evidence to reject it
    as providing a poor fit. \label{f-postcheck}}
  \end{center} 
\end{figure}

  In Figure \ref{f-postcheck} we show histograms for the true observed
  distributions of $z$ and $\log L$. These histograms are compared
  with the posterior median of the distributions based on the mixture
  of Gaussian functions model, as well as error bars containing $90\%$
  of the simulated values. Also shown is a plot comparing the true
  values of the maximum of $L_{obs}$ as a function of $z$ with those
  based on $L^t_{obs}$ and $z^t_{obs}$. As can be seen, the
  distributions of the observed data assuming the mixture of Gaussian
  functions model are consistent with the true distributions of the
  observed data, and therefore there is no reason to reject the
  mixture model as providing a poor fit.

  \section{SUMMARY}
  
  \label{s-summary}

  We have derived the observed data likelihood function which relates
  the quasar LF to the observed distribution of redshifts,
  luminosities. This likelihood function is then used in a Bayesian
  approach to estimating the LF, where the LF is approximated as a
  mixture of Gaussian functions. Because much of this work was
  mathematically technical, we summarize the important points here.
  \begin{itemize}
  \item
    Equation \ref{eq-obslik1} gives the likelihood function for an
    assumed parametric luminosity function. This likelihood function
    differs from the Poisson likelihood commonly used in the LF
    literature because it correctly models the sample size as a
    binomial random variable, whereas the Poisson likelihood
    approximates the sample size as a Poisson random variable. In
    practice, the difference in the maximum-likelihood estimates
    obtained from the two likelihood functions do not seem to be
    significantly different so long as the probability of including a
    source in a survey is small.
  \item
    The product of Equations (\ref{eq-thetapost}) and (\ref{eq-npost})
    is the joint posterior probability distribution of the LF, given
    the observed data. These equations may be used to perform Bayesian
    inference on the LF, after assuming a prior distribution on the LF
    parameters. Bayesian inference is often most easily performed by
    simulating random variables drawn from the posterior probability
    distribution. These random draws may be used to estimate the
    posterior distribution for the LF, as well as to estimate the
    posterior distribution for any quantities calculated from the
    LF. The posterior distribution provides statistically accurate
    uncertainties on the LF and related quantities, even when the
    sample size is small and one is including information below the
    survey detection limits. In contrast, confidence intervals derived
    from bootstrapping the maximum-likelihood estimate can be too
    small.
  \item
    We describe a flexible model for the LF, where the LF is modeled
    as a mixture of Gaussian functions. Equation (\ref{eq-mixmod})
    describes the probability distribution of $\log L$ and $\log z$
    under the mixture of Gaussian functions model, and
    (\ref{eq-mixlf}) describes the LF under the mixture of Gaussian
    functions model.

    Equation (\ref{eq-prior}) gives our prior distribution for the
    Gaussian function parameters. The marginal posterior distribution
    of the mixture model parameters is given by Equation
    (\ref{eq-thetapost_log}), the conditional posterior distribution
    of $N$ at a given $\theta$ is given by Equation (\ref{eq-npost}),
    and the complete joint posterior distribution is the product of
    Equations (\ref{eq-thetapost_log}) and (\ref{eq-npost}).
  \item
    We describe in \S~\ref{s-mha} a Metropolis-Hastings algorithm for
    obtaining random draws from the posterior distribution for the LF
    assuming a Schechter function or mixture of Gaussian functions
    model. In \S~\ref{s-sim}, we use a simulated sample, modeled after
    the SDSS DR3 Quasar Catalog, to illustrate the effectiveness of
    our statistical method, as well as to give an example on how to
    use the Metropolis-Hastings output to perform statistical
    inference on the LF and assess the LF fit.
  \end{itemize}

  So long as the mixture of Gaussian functions model is an accurate
  approximation to the true LF over all luminosities, the
  uncertainties on the LF, assuming the Gaussian mixture model, are
  trustworthy because they are calculated directly from the
  probability distribution of the LF, given the observed
  data. Statistical inference on the LF below the flux limit can
  become significantly biased if one assumes an incorrect and
  restrictive parametric form, as extrapolation errors can become
  large. In this case, the derived posterior may not contain the true
  LF because the wrong parametric model was assumed; this type of
  error is known as model misspecification. For example, consider a
  case when the true LF is a Schechter function, and one is only able
  to detect sources brighter than $L^*$. If one were to assume a
  power-law for the LF, extrapolation below the flux limit would be
  significantly biased. In contrast, the mixture of Gaussian functions
  model, while incorrect, is flexible enough to accurately approximate
  the true Schechter function form, thus minimizing extrapolation bias
  due to model misspecification. Of course, in this example, the most
  accurate results would be obtained by fitting a Schechter function
  to the data, since it is the correct parametric form. Therefore,
  the mixture of Gaussian functions model will not perform as well as
  assuming the correct parametric model, or at least as well as an
  alternative parametric model that better approximates the true LF.

  Although we have focused on the mixture of Gaussian functions model,
  the likelihood and posterior distribution are applicable for any
  parametric form, as illustrated in \S~\ref{s-schechter}. The
  observed data likelihood function for the LF is given by Equation
  (\ref{eq-obslik1}), and the posterior distribution is given by the
  product of Equations (\ref{eq-thetapost}) and
  (\ref{eq-npost}). Then, one can use Equation (\ref{eq-phiconvert})
  to `plug-in' any parametric form of the LF into the appropriate
  likelihood function and posterior distribution, as was done in
  Equation (\ref{eq-schechpost}) for a Schechter function. In
  addition, the metropolis-hastings algorithm is a general method for
  obtaining random draws from the posterior, and can be developed for
  any parametric form of the LF. This therefore allows one to perform
  Bayesian inference for any variety of parametric models of the LF,
  and one is not merely limited to the mixture of Gaussian functions
  model or Schechter function considered in this work.

  An IDL computer routine written for performing the
  Metropolis-Hastings algorithm for the mixture of Gaussian functions
  model is available on request from B. Kelly.

  \acknowledgements

  We acknowledge support from NSF grant AST 03-07384 and a David and
  Lucile Packard Fellowship in Science and Engineering.

  \appendix
  
  \section{MAXIMUM-LIKELIHOOD VS BAYESIAN INFERENCE}

  \label{a-mle_vs_bayes}

  In this section we compare the maximum likelihood approach with the
  Bayesian approach. We do this for readers who are unfamiliar with
  some of the more technical aspects of the two approaches, with the
  hope that the discussion in this section will facilitate
  interpretation of our results in the main body of the paper.

  In maximum-likelihood analysis, one is interested in finding the
  estimate that maximizes the likelihood function of the data. For a
  given statistical model, parameterized by $\theta$, the likelihood
  function, $p(x|\theta)$, is the probability of observing the data,
  denoted by $x$, as a function of the parameters $\theta$. The
  maximum-likelihood estimate, denoted by $\hat{\theta}$, is the value
  of $\theta$ that maximizes $p(x|\theta)$. Under certain regularity
  conditions, $\hat{\theta}$ enjoys a number of useful properties. In
  particular, as the sample size becomes infinite, $\hat{\theta}$
  becomes an unbiased estimate of $\theta$. An unbiased estimator is
  an estimator with expectation equal to the true value, i.e.,
  $E(\hat{\theta}) = \theta_0$, where $\theta_0$ is the true value of
  $\theta$. Therefore, on average, an unbiased estimator will give the
  true value of the parameter to be estimated.

  Because the maximum likelihood estimate is a function of the data,
  $\hat{\theta}$ has a sampling distribution. The sampling
  distribution of $\hat{\theta}$ is the distribution of $\hat{\theta}$
  under repeated sampling from the probability distribution of the
  data.  Under certain regularity conditions, the sampling
  distribution of $\hat{\theta}$ is asymptotically normal with
  covariance matrix equal to the average value of the inverse of the
  Fisher information matrix, $I(\theta)$, evaluated at $\theta_0$. The
  Fisher information matrix is the expected value of the matrix of
  second derivatives of the log-likelihood, multiplied by
  $-1$. Formally, this result states that as $n \rightarrow \infty$,
  then
  \begin{eqnarray}
    \hat{\theta} & \sim & N_p(\theta_0, I^{-1}(\theta_0)), \label{eq-mle_asympt} \\
    I(\theta) & = & -E\left(\frac{\partial^2}{\partial \theta^2} \ln p(x|\theta)\right), 
      \label{eq-fishinf}
  \end{eqnarray}
  where $p$ is the number of parameters in the model, and the
  expectation in Equation (\ref{eq-fishinf}) is taken with respect to
  the sampling distribution of $x$, $p(x|\theta_0)$. Because we do not
  know $\theta_0$, it is common to estimate $I^{-1}(\theta_0)$ by
  $I^{-1}(\hat{\theta})$. In addition, it is common to estimate
  $I(\theta)$ as the matrix of second derivatives of the
  log-likelihood of one's data, since the sample average is a
  consistent estimate for the expectation value. Qualitatively,
  Equation(\ref{eq-mle_asympt}) states that as the sample size becomes
  large, $\hat{\theta}$ approximately follows a normal distribution
  with mean $\theta_0$ and covariance matrix
  $I^{-1}(\hat{\theta})$. This fact may be used to construct
  confidence intervals for $\theta$.

  While the asymptotic results are useful, it is not always clear how
  large of a sample is needed until Equation (\ref{eq-mle_asympt}) is
  approximately true. The maximum likelihood estimate can be slow to
  converge for models with many parameters, or if most of the data is
  missing. Within the context of luminosity function estimation, the
  maximum-likelihood estimate will be slower to converge for surveys
  with shallower flux limits. In addition, Equation
  (\ref{eq-mle_asympt}) does not hold if the regularity conditions are
  not met. In general, this is not a concern, but it is worth noting
  that the asymptotics do not hold if the true value of $\theta$ lies
  on the boundary of the parameter space. For example, in the case of
  a Schechter luminosity function, if the true value of the shape
  parameter, $\alpha$ (see [\ref{eq-schechter}]), is $\alpha_0 = -1$,
  then Equation (\ref{eq-mle_asympt}) does not hold, since $\alpha >
  -1$. If $\alpha_0 \approx -1$, then Equation (\ref{eq-mle_asympt})
  is still valid, but it will take a large sample before the
  asymptotics are valid, as $\alpha_0$ lies near the boundary of the
  parameter space.

  In Bayesian analysis, one attempts to estimate the probability
  distribution of the model parameters, $\theta$, given the observed
  data $x$. The probability distribution of $\theta$ given $x$ is
  related to the likelihood function as
  \begin{equation}
    p(\theta|x) \propto p(x|\theta) p(\theta).
    \label{eq-bayes}
  \end{equation}
  The term $p(x|\theta)$ is the likelihood function of the data, and
  the term $p(\theta)$ is the prior probability distribution of
  $\theta$; the result, $p(\theta|x)$ is called the posterior
  distribution. The prior distribution, $p(\theta)$, should convey
  information known prior to the analysis. In general, the prior
  distribution should be constructed to ensure that the posterior
  distribution integrates to one, but to not have a significant effect
  on the posterior. In particular, the posterior distribution should
  not be sensitive to the choice of prior distribution, unless the
  prior distribution is constructed with the purpose of placing
  constraints on the posterior distribution that are not conveyed by
  the data. The contribution of the prior to $p(\theta|x)$ becomes
  negligible as the sample size becomes large.

  From a practical standpoint, the primary difference between the
  maximum likelihood approach and the Bayesian approach is that the
  maximum likelihood approach is concerned with calculating a point
  estimate of $\theta$, while the Bayesian approach is concerned with
  mapping out the distribution of $\theta$. The maximum likelihood
  approach uses an estimate of the sampling distribution of
  $\hat{\theta}$ to place constraints on the true value of
  $\theta$. In contrast, the Bayesian approach directly calculates the
  probability distribution of $\theta$, given the observed data, to
  place constraints on the true value of $\theta$. It is illustrative
  to consider the case when the prior is taken to be uniform over
  $\theta$; assuming the posterior integrates to one, the posterior is
  then proportional to the likelihood function, $p(\theta|x) \propto
  p(x|\theta)$. In this case, the goal of maximum likelihood is to
  calculate an estimate of $\theta$, where the estimate is the most
  probable value of $\theta$, given the observed data. Then,
  confidence intervals on $\theta$ are derived from the maximum
  likelihood estimate, $\hat{\theta}$, usually by assuming Equation
  (\ref{eq-mle_asympt}). In contrast, the Bayesian approach is not
  concerned with optimizing the likelihood function, but rather is
  concerned with mapping out the likelihood function. Under the
  Bayesian approach with a uniform prior, confidence intervals on
  $\theta$ are derived directly from likelihood function, and an
  estimate of $\theta$ can be defined as, for example, the value of
  $\theta$ averaged over the likelihood function. So, the maximum
  likelihood attempts to obtain the `most likely' value of $\theta$,
  while the Bayesian approach attempts to directly obtain the
  probability distribution of $\theta$, given the observed
  data. Because the Bayesian approach directly estimates the
  probability distribution of $\theta$, and because it does not rely
  on any asymptotic results, we consider the Bayesian approach to be
  preferable for most astronomical applications.

  \section{DERIVATION OF THE MARGINAL POSTERIOR DISTRIBUTION FOR TRUNCATED DATA}

  \label{a-margpost_deriv}

  Here, we give a derivation of the posterior probability distribution
  of $\theta$, given by Equation (\ref{eq-thetapost}). If we assume a
  uniform prior on $\log N$, then this is equivalent to assuming the
  prior $p(\theta,N) \propto N^{-1} p(\theta)$. In this case, the
  posterior distribution is given by
  \begin{equation}
    p(\theta, N | L_{obs},z_{obs}) \propto N^{-1} p(\theta) C^N_n \left[
    p(I=0|\theta) \right]^{N-n} \prod_{i \in {\cal A}_{obs}} p(L_i,z_i | \theta).
  \end{equation}
  The marginal posterior distribution of $\theta$ is obtained by
  summing the joint posterior over all possible values of $N$. For the
  choice of prior $p(\theta,\log N) \propto p(\theta)$, the marginal
  posterior of $\theta$ is
  \begin{eqnarray}
    p(\theta|L_{obs},z_{obs}) & \propto & p(\theta) \left[\prod_{i \in {\cal A}_{obs}} 
      p(L_i, z_i|\theta) \right] \sum_{N=n}^{\infty} N^{-1} C^N_n  \left[ p(I=0|\theta) \right]^{N-n} 
    \label{eq-thetamarg1} \\
    & \propto & p(\theta) \left[ p(I=1|\theta) \right]^{-n} \left[\prod_{i \in {\cal A}_{obs}} 
      p(L_i, z_i|\theta) \right] \sum_{N=n}^{\infty} C^{N-1}_{n-1} \left[ p(I=0|\theta) \right]^{N-n}
    \left[ p(I=1|\theta) \right]^{n}, \label{eq-thetamarg2}
  \end{eqnarray}
  where we arrived at the second Equation by multiplying and dividing
  the first Equation by $p(I=1|\theta)^n$ and noting that $C^N_n =
  C^{N-1}_{n-1} (N / n)$. The term within the sum is the mathematical
  expression for a negative binomial distribution as a function of $N$
  (see Eq.[\ref{eq-negbin}]). Because probability distributions must
  be equal to unity when summed over all possible values, the sum is
  just equal to one. We therefore arrive at Equation
  (\ref{eq-thetapost}) by replacing the summation in Equation
  (\ref{eq-thetamarg2}) with the value of one.

  \section{SOME PROBABILITY DISTRIBUTIONS USED IN THIS WORK}

  \label{a-densities}

  In this section of the appendix we briefly describe some probability
  distribution that we employ, but may be unfamiliar to some
  astronomers.

  \subsection{Negative Binomial}

  The negative binomial distribution is closely related to the
  binomial distribution. The binomial distribution gives the
  probability of observing $n$ `successes', given that there have been
  $N$ trials and that the probability of success is $p$. In contrast,
  the negative binomial distribution gives the probability of needing
  $N$ trials before observing $n$ successes, given that the
  probability of success is $p$. Within the context of this work, the
  binomial distribution gives the probability of detecting $n$
  sources, given that there are $N$ total sources and that the
  detection probability is $p$. The negative binomial distribution
  gives the probability that the total number of sources is $N$, given
  that we have detected $n$ sources and that the detection probability
  is $p$. The negative binomial distribution is given by
  \begin{equation}
    p(N|n,p) = C^{N-1}_{n-1} p^n (1 - p)^{N-n}, \ \ N \geq n.
    \label{eq-negbin}
  \end{equation}
  
  A random draw from the negative binomial distribution with
  parameters $n$ and $p$ may be simulated by first drawing $n$ random
  values uniformly distributed on $[0,1]$, $u_1, \ldots, u_n \sim {\rm
  Uniform}(0,1)$. Then, calculate the quantity
  \begin{equation}
    m = \sum_{i=1}^{n} \left\lfloor \frac{\log u_i}{\log (1 - p)} \right\rfloor, \label{eq-negbin_sim}
  \end{equation}
  where $\lfloor \cdot \rfloor$ is the floor function, i.e., $\lfloor
  x \rfloor$ denotes the greatest integer less than or equal to
  $x$. The quantity $N = n + m$ will then follow a negative binomial
  distribution with parameters $n$ and $p$.

  \subsection{Dirichlet}

  The Dirichlet distribution is a multivariate generalization of the
  Beta distribution, and it is commonly used when modeling group
  proportions. Dirichlet random variables are constrained to be
  positive and sum to one. The Dirichlet distribution with argument
  $\theta_1,\ldots,\theta_k$ and parameters $\alpha_1, \ldots, \alpha_k$
  is given by
  \begin{equation}
    p(\theta_1,\ldots,\theta_k|\alpha_1,\ldots,\alpha_k) = 
    \frac{\Gamma(\alpha_1 + \ldots + \alpha_k)}{\Gamma{\alpha_1}\cdots\Gamma{\alpha_k}}
    \prod_{i=1}^{k} \theta_i^{\alpha_i-1}, \ \ \theta_1, \ldots, \theta_k \geq 0, \ 
    \alpha_1, \ldots, \alpha_k > 0,\ \ \sum_{i=1}^k \theta_i = 1. \label{eq-dirichlet}
  \end{equation}
  To draw a random value $\theta_1, \ldots, \theta_k$ from a Dirichlet
  distribution with parameters $\alpha_1, \ldots, \alpha_k$, first
  draw $x_1, \ldots, x_k$ independently from Gamma distributions with
  shape parameters $\alpha_1, \ldots, \alpha_k$ and common scale
  parameter equal to one. Then, set $\theta_j = x_j / \sum_{i=1}^k
  x_i$. The set of $\theta$ will then follow a Dirichlet distribution. 

  \subsection{Multivariate Student-$t$ and Cauchy Distribution}

  The Student-$t$ distribution is often used as a robust alternative
  to the normal distribution because it is more heavily tailed than
  the normal distribution, and therefore reduces the effect of
  outliers on statistical analysis. A $t$ distribution with $\nu = 1$
  degree of freedom is referred to as a Cauchy distribution, and it is
  functionally equivalent to a Lorentzian function. A $p$-dimensional
  multivariate $t$ distribution with $p$-dimensional argument ${\bf
  x}$, $p$-dimensional mean vector $\mu$, $p \times p$ scale matrix
  $\Sigma$, and degrees of freedom $\nu$ is given by
  \begin{equation}
    p({\bf x}|\mu, \Sigma, \nu) = \frac{\Gamma((\nu + p)/2)}{\Gamma(\nu/2) 
      \nu^{p/2} \pi^{p/2}} |\Sigma|^{-1/2} \left[1 + \frac{1}{\nu} \left({\bf x} - \mu\right)^T
      \Sigma^{-1} \left({\bf x} - \mu \right) \right]^{-(\nu + p)/2}.
    \label{eq-tdist}
  \end{equation}
  The 1-dimensional $t$ distribution is obtained by replacing matrix
  and vector operations in Equation (\ref{eq-tdist}) with scalar
  operations. 

  Although we do not simulate from a $t$ distribution in this work,
  for completeness we include how to do so. To simulate a random
  vector ${\bf t}$ from a multivariate $t$ distribution with mean
  vector $\mu$, scale matrix $\Sigma$, and degrees of freedom $\nu$,
  first draw ${\bf z}$ from a zero mean multivariate normal
  distribution with covariance matrix $\Sigma$. Then, draw $x$ from a
  chi-square distribution with $\nu$ degrees of freedom, and compute
  the quantity ${\bf t} = \mu + {\bf z} \sqrt{\nu / x}$. The quantity
  ${\bf t}$ is then distributed according to the multivariate $t$
  distribution.

  \subsection{Wishart and Inverse Wishart}

  The Wishart distribution describes the distribution of the $p \times
  p$ sample covariance matrix, given the $p \times p$ population
  covariance matrix, for data drawn from a multivariate normal
  distribution. Conversely, the inverse Wishart distribution describes
  the distribution of the population covariance matrix, given the
  sample covariance matrix, when the data are drawn from a
  multivariate normal distribution. The Wishart distribution can be
  thought of as a multivariate extension of the $\chi^2$
  distribution. A Wishart distribution with $p \times p$ argument $S$,
  $p \times p$ scale matrix $\Sigma$, and degrees of freedom $\nu$ is
  given by
  \begin{equation}
    p(S|\Sigma,\nu) = \left[ 2^{\nu p / 2} \pi^{p (p-1)/4} \prod_{i=1}^{p} 
      \Gamma \left(\frac{\nu + 1 - i}{2} \right) \right]^{-1}
    |\Sigma|^{-\nu/2} |S|^{(\nu - p - 1) / 2} \exp\left \{ -\frac{1}{2} 
    tr (\Sigma^{-1} S) \right \} , \label{eq-wishart}
  \end{equation}
  where the matrices $S$ and $\Sigma$ are constrained to be positive
  definite. An inverse Wishart distribution with $p \times p$ argument
  $\Sigma$, $p \times p$ scale matrix $S$, and degrees of freedom
  $\nu$ is
  \begin{equation}
    p(\Sigma|S,\nu) = \left[ 2^{\nu p / 2} \pi^{p (p-1)/4} \prod_{i=1}^{p} 
      \Gamma \left(\frac{\nu + 1 - i}{2} \right) \right]^{-1}
    |S|^{\nu/2} |\Sigma|^{-(\nu + p + 1) / 2} \exp\left\{-\frac{1}{2} tr(\Sigma^{-1} S)\right\},
    \label{eq-invwishart}
  \end{equation}
  where the matrices $S$ and $\Sigma$ are constrained to be positive
  definite.

  To draw a $p \times p$ random matrix from a Wishart distribution
  with scale matrix $\Sigma$ and $\nu$ degrees of freedom, first draw
  ${\bf x}_1, \ldots, {\bf x}_{\nu}$ from a zero mean multivariate
  normal distribution with $p \times p$ covariance matrix
  $\Sigma$. Then, calculate the sum $S = \sum_{i=1}^{\nu} {\bf x}_i
  {\bf x}_i^T$. The quantity $S$ is then a random draw from a Wishart
  distribution. Note that this technique only works when $\nu \geq
  p$. A random draw from the inverse Wishart distribution with scale
  matrix $S$ and degrees of freedom $\nu$ may be obtained by first
  obtaining a random draw $W$ from a Wishart distribution with scale
  matrix $S^{-1}$ and degrees of freedom $\nu$. The quantity $\Sigma =
  W^{-1}$ will then follow an inverse Wishart distribution.

\end{document}